\documentclass[useAMS,usenatbib]{mn2e}

\usepackage{amssymb}
\usepackage{amsfonts}
\usepackage{natbib}
\usepackage{epsfig}
\usepackage{psfig}
\usepackage{mncite}

\newcommand{\de}{\mbox{d}}

\newcommand{\msunyr}{M_{\odot}/\mbox{yr}}
\newcommand{\mdot}{\dot{M}}

\begin{document}

\title[Black hole formation in pre-galactic discs]{Supermassive black
hole formation during the assembly of pre-galactic discs}

\author[Lodato \&  Natarajan] {Giuseppe Lodato$^1$ and Priyamvada 
Natarajan$^{2,3}$\\
$^1$ Institute of Astronomy, Madingley Road, Cambridge, CB3 0HA\\
$^2$ Department of Astronomy, Yale University, P. O. Box 208101, 
New Haven, CT 06511-208101, USA \\
$^3$ Department of Physics, Yale University, P. O. Box 208120, 
New Haven, CT 06520-208120, USA}

\maketitle

\begin{abstract}
  In this paper we discuss the evolution of gravitationally unstable
  pre-galactic discs that result from the collapse of haloes at high
  redshift $z \approx 10$ or so, which have not yet been enriched by
  metals. In cases where molecular hydrogen formation is suppressed
  the discs are maintained at a temperature of a few thousand degrees
  Kelvin.  However, when molecular hydrogen is present cooling can
  proceed down to a few hundred degrees Kelvin. Analogous to the case
  of the larger scale proto-galactic discs, we assume that the
  evolution of these discs is mainly driven by angular momentum
  redistribution induced by the development of gravitational
  instabilities in the disc. We also properly take into account the
  possibility of disc fragmentation. We thus show that this simple
  model naturally predicts the formation of supermassive black holes
  in the nuclei of such discs and provides a robust determination of
  their mass distribution as a function of halo properties. We
  estimate that roughly 5\% of discs resulting from the collapse of
  haloes with $M\approx 10^7 M_{\odot}$ should host a massive black
  hole with a mass $M_{\rm BH}\approx 10^5 M_{\odot}$. We confirm our
  arguments with time-dependent calculations of the evolution of the
  surface density and of the accretion rate in these primordial
  discs. The luminosity of the outer, colder disc is expected to be
  very low (in the range of a few thousand $L_{\odot}$), while the
  formation of the black hole is expected to produce a burst with a
  luminosity of a few times $10^9L_{\odot}$. This mechanism offers an
  efficient way to form seed black holes at high redshift. The
  predicted masses for our black hole seeds enable the comfortable
  assembly of $10^9 M_{\odot}$ black holes powering the luminous
  quasars detected by the Sloan Digital Sky Survey at $z = 6$ for a
  concordance cosmology.
\end{abstract}

\section{Introduction}

The local demography of black holes at the centers of galaxies suggests
that black hole formation is a generic feature of galaxy formation
\citep{magorrian98,tremaine02,ferrarese2000}. Mergers and accretion
processes, in fact, merger induced accretion are implicated in the
assembly of black holes \citep{kauffmann00,volonteri03,dimatteo05}.
However, all models require the formation of seed black holes at high
redshift. The inferred large masses ($M_{\rm bh} \sim 10^9 M_{\odot}$)
of the black holes powering luminous quasars at $z \sim 6$ detected by
the Sloan Digital Sky Survey (SDSS) offers a challenge to mechanisms
for production of seed black holes at higher redshifts
\citep{fan04}. In a $\Lambda$CDM cosmology, these black holes have to
assemble such large masses within a Gyr. The popular picture is that
the remnants of the first generation of metal-free stars that form with
an initial mass function that is biased towards higher masses
\citep{abel00,bromm02} provide the initial seed black holes with masses
of the order of a 10's to 100 $M_{\odot}$
\citep{madaurees01,volonteri03,ricotti04,mapelli06}. Growing these
seeds to the requisite masses by $z = 6$ to match the abundance of
observed SDSS quasars requires various kinds of fine-tuning including
phases of maximal growth powered by super-Eddington accretion.
\citet{volonteri05b} consider the growth of black holes in metal-free
halos with $T > 10^4$ K that cool via atomic hydrogen lines leading to
the formation of a fat, gas disc. They argue that an episode of
super-Eddington accretion assembles black holes with masses of
$10^6\,M_{\odot}$ at $z \sim 15 - 20$.

There are alternative models that predict the direct formation of more
massive seeds with masses of about $10^5\,M_{\odot}$.  These range from
scenarios based on the formation of supermassive objects formed
directly out of the collapse of dense gas
\citep{haehnelt93,umemura93,loeb94,eisenstein95,bromm03,koushiappas04,begelman06}.
The key limiting factor for these models is the disposal of the angular
momentum. One approach taken by \citet{eisenstein95} and
\citet{koushiappas04} is to argue that it is preferentially either low
spin halos with consequently low angular momentum gas or the low spin
tail of the gas distribution in halos that can cool efficiently that
form these seeds. Significant transfer of angular momentum is still
required to go from the central massive object to the final collapsed
hole that is expected to occur via the post-Newtonian
instability. Non-axisymmetric structures like bars have been proposed
\citep{shlosman90} to efficiently transfer angular momentum at these
late stages. In a recent paper, \citet{begelman06} have pursued this
picture and argue that a low specific entropy `quasi-star' is produced
as a result of the cascade of the bars-within-bars instability leading
to the formation of a black hole of a few solar masses. 

In this paper, we argue for the formation of more massive black hole
seeds ($\sim 10^5\,M_{\odot}$) from gas cooling in primordial halos via
the growth of gravitational instabilities. We offer a picture wherein
accretion processes and fragmentation criteria are addressed in a
coupled fashion. The outline of our paper is as follows: in Section~2,
we outline the scenario for pre-galactic disc formation in dark matter
halos and discuss their accretion properties, the fragmentation
criteria which determine the fate of the gas are discussed in Section~3
and the implied mass distribution of central concentrations is derived
in Section~4. Detailed time-dependent models that confirm the analytic
calculations in the previous sections are presented in Section~5, the
resultant luminosity and potential observability of these sources is
discussed in Section~6 followed by a discussion and implications of our
work.

\section{Pre-galactic disc formation}

In the context of the hierarchical, cold dark matter dominated
structure formation paradigm, the baryonic components of galaxies
assemble upon dissipation within dark matter halos. Generic to these
models is the assembly of discs in these haloes as a consequence of
dissipational collapse of the baryons
\citep{white78,fall80,mo98,dalcanton97}. Here, we revisit some of the
standard arguments put forward to describe the formation of
proto-galactic discs inside a spherical dark matter halo of mass
$M$. In particular, we adopt the formalism developed by \citet{mo98},
and extend it to include the effect of angular momentum redistribution
and mass accretion due to gravitational instabilities in the gas disc.

The arguments that form the basis of the standard disc formation
picture are simple and elegant. It is assumed that a fraction $m_{\rm
d}$ of the total mass of the halo, containing a fraction $j_{\rm d}$ of
the angular momentum of the halo (usually assumed to be equal to
$m_{\rm d}$), collapses and forms a disc. Note that the baryonic mass
in the halo is assumed to be a fixed fraction of the halo mass, and the
proportionality factor is taken to be the universal baryon fraction
$f_b$.  Assuming a radial profile for the resultant gas density
distribution in the disc (for example, exponential) enables the
calculation of its key physical properties: the disc scale length,
central surface mass density, and crucially its stability under
self-gravity. Star formation is then assumed to occur for those discs
that are gravitationally unstable. The results of earlier work based on
this paradigm depend on the detailed assumptions made. For example,
while \citet{mo98} consider both an isothermal sphere and a NFW
\citep{nfw} profile for the dark matter halo and a global criterion for
gravitational instability, \citet{oh} consider an NFW \citep{nfw} halo
and a local instability criterion, based on the Toomre-$Q$
parameter. These treatments present `static' models for disc
assembly. In this work, we examine the fate of these pre-galactic discs
taking their time dependent evolution into account.

The limitation of previous analyses is that they consider the
properties of the discs {\it a posteriori}, after all of their mass has
been assembled, and do not consider the evolution of the discs {\it
during} their formation process. They have argued that for some values
of the relevant halo parameters, in particular, for low values of the
spin parameter of the halo $\lambda$ \footnote{The spin parameter
$\lambda$ of a dark matter halo is defined as $\lambda =
J\,|E|^{1/2}/G\,M^{5/2}$, where $J$ is the angular momentum of the
halo, $E$ its total energy and $M$ the halo mass.}, the disc can be
self-gravitating. In principle, these analyses also allow the disc to
have very low values of $Q$ or equivalently to be strongly unstable in
terms of a stability criterion. However, it is important to note here
that as soon as the disc becomes massive enough to be marginally
stable, it will develop structures that will re-distribute angular
momentum and mass through the disc, preventing the surface density from
becoming too large and the disc from becoming too unstable. A
suggestion that this kind of matter redistribution in primordial discs
can lead to the formation of seed black holes was made recently by
\citet{begelman06} and also, previously, by \citet{koushiappas04} but
using substantially different arguments compared to our work here. In
practice, the accretion of matter to the center will stop as soon as
the surface density becomes low enough to make the disc stable, so that
in the final state the disc will be exactly marginally stable. We
explore this scenario and the consequences of angular momentum
redistribution in detail here. In this section, we propose some simple
arguments that allow us to estimate the final disc mass and the mass
that will flow into the central region, leading to the formation and
growth of a seed black hole. In Section \ref{sec:time} we support these
simple arguments with full time-dependent calculations. We are
therefore able to keep a detailed, self-consistent inventory of the
fate of the gas.

We assume that in the final configuration, once the pre-galactic disc
has assembled in the dark matter halo, only a fraction $m_{\rm f}$ of
the total halo mass is retained in the disc (note that we track the
baryons that remain in the disc and those that are accreted towards the
center separately), while the remainder $m_{\rm a}=m_{\rm d}-m_{\rm f}$
is accreted towards the center of the galaxy providing fuel for a
growing black hole. The final angular momentum per unit mass of the
disc, however, remains equal to the initial value $j_{\rm d}$, since,
in any accretion process, angular momentum is transported outwards, and
the only loss of angular momentum due to accretion is related to the
advection of it out of the inner radius of the disc. For an isothermal
dark matter halo $\rho(r) \propto 1/r^2$ with a flat rotation curve,
the angular momentum advected inwards due to the accretion of a mass
$m_{\rm a}$ is $j_{\rm a}=j_{\rm d}R_{\rm in}/R_{\rm d}<<j_{\rm d}$
(here $R_{\rm in}$ and $R_{\rm d}$ are the inner and outer radius of
the disc, respectively). It is assumed here that the disc is embedded
in a spherical, isothermal, dark matter halo, with virial temperature
$T_{\rm vir}$ and a constant rotational velocity $V_{\rm h}$. To
evaluate the stability of the disc, we use the Toomre parameter $Q$
which is defined as:
\begin{equation}
Q=\frac{c_{\rm s}\kappa}{\pi G \Sigma}=\sqrt{2}\frac{c_{\rm s}V_{\rm
h}}{\pi G\Sigma R},
\label{Q}
\end{equation}
where $c_{\rm s}$ is the sound speed, $\kappa=\sqrt{2}V_{\rm h}/R$ is
the epicyclic frequency, $R$ is the cylindrical radial coordinate, and
$\Sigma$ is the surface mass density of the disc. We consider here the
earliest generations of discs, which have not been metal-enriched and
therefore are able to cool only through hydrogen. We present in the
Appendix the detailed structure of these discs. Here we note
that, in thermal equilibrium, if molecular hydrogen formation is {\it
inhibited}, these discs are expected to be nearly isothermal at a
temperature of a few thousand K. If molecular hydrogen is
present, however, the disc can become much colder, with a temperature
of a few hundred K. A marginally stable, isothermal disc
has the same surface density profile as a Mestel disc \citep{mestel63}.
We therefore assume that:
\begin{equation}
\Sigma(R)=\Sigma_0\left(\frac{R}{R_{\rm d}}\right)^{-1}.
\end{equation}
Dark matter halos are characterised by their mass $M$ and their angular
momentum $J$, or equivalently by their spin parameter $\lambda$. The
distribution of spin parameters for cold dark matter halos as determined
from cosmological N-body simulations is given by:
\begin{equation}
p(\lambda)\de \lambda=\frac{1}{\sqrt{2\pi}\sigma_{\lambda}}
\exp\left[-\frac{\ln^2(\lambda/\bar{\lambda})}{2\sigma_{\lambda}}\right]
\frac{\de\lambda}{\lambda},
\label{eq:lambda}
\end{equation}
where the mean spin $\bar{\lambda}=0.05$ and the dispersion is
$\sigma_{\lambda}=0.5$ \citep{warren92}. Requiring that the mass of the
gas disc is a fraction ($m_{\rm d}-m_{\rm a}$) of the halo mass and
that its angular momentum is a fraction $j_{\rm d}$ of the halo,
following \citet{mo98} we easily get:
\begin{equation}
R_{\rm d}=2\sqrt{2}\lambda\left(\frac{j_{\rm d}}{m_{\rm d}}
\right)\frac{1}{1-m_{\rm a}/m_{\rm d}}r_{200},
\label{eq:rd}
\end{equation}
\begin{equation}
\Sigma_0=\frac{10}{16\pi}\frac{m_{\rm d}}{\lambda^2}\left(\frac{m_{\rm
d}}{j_{\rm d}}\right)^2\left(1-\frac{m_{\rm a}}{m_{\rm d}}\right)^3
\frac{H(z)V_{\rm h}}{G},
\end{equation}
where $r_{200}=V_{\rm h}/10H(z)$ is the halo virial radius (the radius
within which the density is 200 times the critical density of the
Universe) and $H(z)$ is the Hubble constant as a function of redshift
$z$. We can now evaluate $Q$, given the disc temperature $T_{\rm gas}$:
\begin{equation}
Q=\frac{8}{m_{\rm d}}\lambda\left(\frac{j_{\rm d}}{m_{\rm d}}\right)
\left(\frac{T_{\rm gas}}{T_{\rm vir}}\right)^{1/2}
\left(\frac{1}{1-m_{\rm a}/m_{\rm d}}\right)^2.
\end{equation}
We use the fact that the final disc configuration is exactly
marginally stable, and thus obtain $m_{\rm a}$ by requiring that
$Q=Q_{\rm c}$, a critical value above which the disc is gravitationally
stable and no accretion can take place, due to the lack of an angular
momentum transport mechanism.  This gives:
\begin{equation}
\frac{m_{\rm a}}{m_{\rm d}}=1-\sqrt{\frac{8\lambda}{m_{\rm d}Q_{\rm c}}
\left(\frac{j_{\rm d}}{m_{\rm d}}\right)\left(\frac{T_{\rm gas}}{T_{\rm
vir}}\right)^{1/2}}.
\label{eq:accreted}
\end{equation}

Note that, although the result summarized in eqn. (\ref{eq:accreted})
is based on the assumption that angular momentum is redistributed
within the disc, it does not depend on the specific viscosity
mechanism, and is therefore very robust. The only assumption made is
that whatever this mechanism is, it is only active when the disc is
gravitationally unstable (a very reasonable assumption for such
primordial discs, for which the main source of viscosity comes from
gravitational instabilities). As shown, the amount of mass that will
be concentrated in the central regions of these pre-galactic discs
depends only on halo properties (such as the spin parameter $\lambda$
and the fraction of baryonic mass that collapses to the disc $m_{\rm
d}$), the ratio between gas temperature and halo virial temperature,
and on the threshold value of $Q$, which has a very small range of
variation around $Q_{\rm c}\approx 1$.

Equation (\ref{eq:accreted}) then provides a powerful link between the
properties of dark matter haloes and the mass of massive, seed
black holes that can grow within them. This implies a larger mass
concentration in the center for haloes with low spin parameter
$\lambda$ or with high virial temperature $T_{\rm vir}$ (corresponding
to higher halo mass mass $M$). The fact that seed black holes are 
likely to form preferentially in low spin halos was also demonstrated by
\citet{eisenstein95} and by \citet{koushiappas04}, using a different
set of arguments.  However, for high mass haloes, there is the added
possibility that the disc fragments. We discuss the issue of
fragmentation in Section \ref{sec:frag} and we postpone to Section
\ref{sec:prob} the discussion of the mass distribution of
central concentrations given the distribution of the spin parameter
$\lambda$.

\section{Fragmentation criteria}
\label{sec:frag}

The arguments outlined in the previous section implicitly assume that,
when the disc is gravitationally unstable, it is able to redistribute
mass and angular momentum so as to accommodate the incoming mass inflow
from the halo, maintaining a marginally stable state. Actually, in some
cases, this is not going to be the case and the disc might undergo
fragmentation. The criterion for the fragmentation of a gravitationally
unstable disc has been studied extensively in the past few years
(especially in relation to models of planet formation). \citet{gammie01}
has shown that being unstable (i.e. having $Q\lesssim 1$) is a
necessary but not sufficient requirement for disc fragmentation. He has
shown that fragmentation occurs only when the cooling time is faster
than the disc dynamical timescale. An equivalent fragmentation
criterion, that directly relates accretion and fragmentation
properties, has been discussed by \citet{RLA05}. Indeed, they have
shown that the gravitationally induced stress in a thin disc cannot
exceed a critical value, which, measured in terms of the standard
$\alpha$ description of transport properties of accretion disc,
corresponds to $\alpha_{\rm c}\approx 0.06$. If the disc is required to
provide a larger stress, by either strong cooling or by a large mass
inflow, it will undergo fragmentation within a dynamical time-scale. In
the context of pre-galactic discs that we are considering here, this
will then lead to efficient star formation in the disc.

In the case considered here, the constraints imposed by cooling are
not important, since, due to the steepness of the cooling function
when the gas is predominantly hydrogen (in either the molecular or
atomic form), slight changes in the temperature can easily bring the
disc into thermal equilibrium. The constraint coming from the inflow
of mass from the halo is much more important. The mass accretion rate
$\dot{M}_{\rm h}$ from the halo can be simply estimated by requiring
that a mass $m_{\rm d}M$ (where $M=V_{\rm h}^2r_{200}/G$ is the total
halo mass) collapses in a free-fall time-scale $t_{\rm
ff}=r_{200}/V_{\rm h}$. We therefore get:
\begin{equation}
\dot{M}_{\rm h}=m_{\rm d}\frac{V_{\rm h}^3}{G}.
\label{eq:mhalo}
\end{equation}

For a self-gravitating disc the standard mass conservation equation for
a steady disc:

\begin{equation}
2\pi\nu\Sigma=\mdot,
\label{eq:cons}
\end{equation}
where $\mdot$ is the mass accretion rate and $\nu$ is the viscosity
(described here with the standard $\alpha$-parameterization), takes the
simple form:

\begin{equation}
\mdot=2\alpha\frac{c_{\rm s}^3}{G},
\label{eq:mdot}
\end{equation}
from which we see that in a steady state any isothermal disc will be
characterized by a constant viscosity coefficient $\alpha$. Note that
in eqn.~(\ref{eq:cons}) there is a factor 2 rather than the standard
factor 3 of Keplerian discs, because the rotation curve here is
assumed to be flat.

What would happen to the transport properties in the disc once the
fragmentation threshold is overcome is still not clear. A plausible
assumption is that the accretion rate would be capped at its threshold
value, with extra heating needed to reach thermal balance. This heat
could be provided by the star formation process. The disc would then
not be able to transfer down to its inner parts an $\mdot$ larger than:

\begin{equation}
\mdot_{\max}=2\alpha_{\rm c}\frac{c_{\rm s}^3}{G},
\end{equation}
where $\alpha_{\rm c}\approx 0.06$. 

If we consider the case of cooling via atomic hydrogen and assume that
$T_{\rm gas}\approx 4000$ K, we find that $\mdot_{\rm max}\approx
10^{-2}\msunyr$.  In contrast, for the cold discs arising when
molecular cooling is efficient, the maximum accretion rate is
significantly lowered, down to $\mdot_{\rm max}\approx 3\times
10^{-4}\msunyr$.
 
If the disc evolves to a steady state, one should expect that the mass
flow in the disc would equal the inflow rate $\dot{M}_{\rm
h}$. Actually, as the disc builds up, only a fraction of the matter
flows into the inner disc, since the remaining has to be transported
outwards to take up the angular momentum lost by in-flowing
material. Since the viscous time-scale in the outer disc is very long,
even if the inner disc is almost in a steady state, the inner flux
$\dot{M}_{\rm in}$ will still be a fraction of $\dot{M}_{\rm h}$, that
we will calculate explicitly in Section \ref{sec:time}. We anticipate
here the result that the mass flux in the inner disc is at most a
fraction $f=1/2(1+m_{\rm a}/m_{\rm d})$ of the total inflow of
matter. Note that $1/2<f<1$, so even in the extreme case where no mass
in accreted (i.e. $m_{\rm a}=0$), the mass flow is only reduced by a
factor of a half.  We expect fragmentation if:
\begin{equation}
\dot{M}_{\rm in}=m_{\rm d}\frac{V_{\rm h}^3}{G}\frac{1}{2}\left(1+
\frac{m_{\rm a}}{m_{\rm d}}\right)>2\alpha_{c}\frac{c_{\rm s}^3}{G}.
\label{eq:frag}
\end{equation}
We can rearrange eqn.~(\ref{eq:frag}) in terms of the ratio of the
virial temperature to the gas temperature:
\begin{equation}
\frac{T_{\rm vir}}{T_{\rm gas}}>\left(\frac{4\alpha_{\rm c}}{m_{\rm
d}}\frac{1}{1+m_{\rm a}/m_{\rm d}}\right)^{2/3}.
\end{equation}
Assuming $\alpha_{\rm c}=0.06$ and $m_{\rm d}=0.05$, we can therefore
identify three possible behaviours:
\begin{itemize}
\item $T_{\rm vir}/T_{\rm gas}\gtrsim 2.9$ : these haloes will form
discs that will fragment and form stars.
\item $1.8\lesssim T_{\rm vir}/T_{\rm gas}\lesssim 2.9$: these haloes
will form fragmenting discs only if the spin parameter is significantly
low, so that $m_{\rm a}/m_{\rm d}\sim O(1)$.
\item $T_{\rm vir}/T_{\rm gas}\lesssim 1.8$: these haloes will not
produce fragmenting discs, but since the halo mass is relatively low,
the amount of mass that can be concentrated in the center
will also be correspondingly low (cf. eqn.~\ref{eq:accreted}).
\end{itemize}
In practice, whenever it happens, fragmentation will act to suppress
the concentration of the largest amount of mass in the center, whereas
haloes with relatively smaller central concentrations are by and large
unaffected by fragmentation.

The criterion for fragmentation described above, and the critical
value of $\alpha$, have been obtained in the context of thin and
Keplerian discs. The haloes that give rise to large concentration of
mass in the center have $T_{\rm vir}\gtrsim T_{\rm gas}$ and therefore
the discs that form within them are relatively thick. In this case,
the Jeans mass in a fragmenting disc is only slightly smaller than the
total disc mass, so that fragmentation might be less likely in this
case than for thin discs, effectively leading to an increase of the
critical value of $\alpha$.

\section{Mass distribution of central concentrations}
\label{sec:prob}

We now discuss the distribution of central mass concentrations, given
a distribution of halo spin parameters. We determine the distribution
of the mass available to feed a growing black hole in the center of
the pre-galactic disc, $M_{\rm BH}=m_{\rm a}M$, based on eqn
(\ref{eq:accreted}), while the distribution of $\lambda$ is given by
eqn. (\ref{eq:lambda}). We also need to specify the values of $m_{\rm
d}$ and $Q_{\rm c}$. We assume $m_{\rm d}=0.05$, but we also consider
the case were $m_{\rm d}=0.1$. As customary, we take $j_{\rm d}=m_{\rm
d}$ \citep{mo98}. The critical value for the Q-parameter, $Q_{\rm c}$
is more uncertain.  A self-gravitating disc is marginally stable to
local, axi-symmetric instabilities at $Q=1$. However, non
axi-symmetric and global instabilities cause the disc to become
unstable at larger values of $Q$. The prescription described above
implies that at $Q=Q_{\rm c}$ the disc provides no stress, while on
the other hand, \citet{LR04,LR05} have shown that a $Q=1$ disc is able
to deliver a sizable stress. We expect $Q_{\rm c}$ to be of
the order of, but relatively larger than, unity and therefore consider
two cases: (i) $Q_{\rm c}=2$ and (ii) $Q_{\rm c}=3$.

\begin{figure*}
\centerline{\epsfig{figure=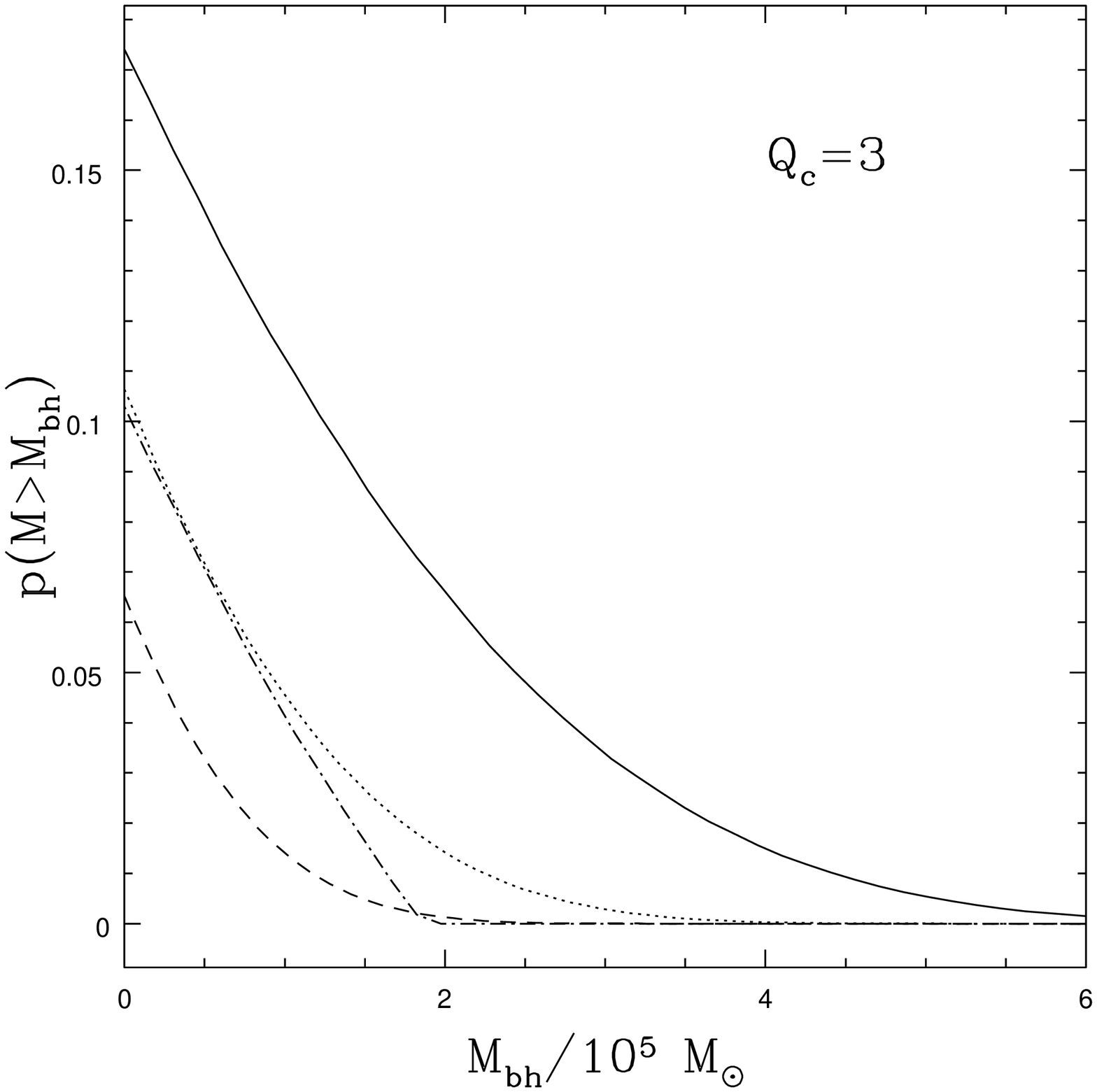,width=0.5\textwidth}
            \epsfig{figure=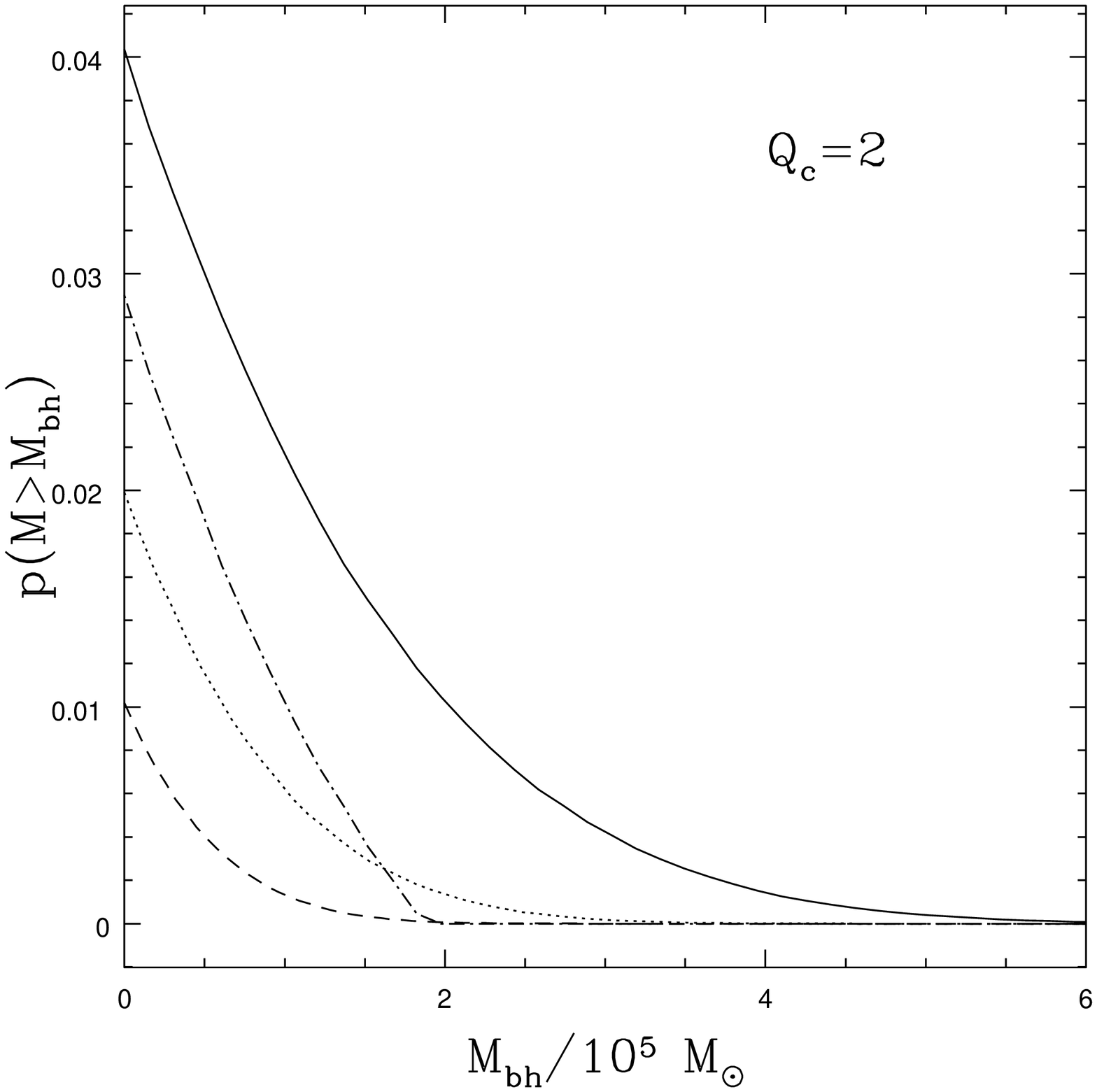,width=0.5\textwidth}}
            \caption{Cumulative distribution of the mass available to
            the formation of a seed black hole: The parent haloes are
            at a redshift $z=10$ and we assume here that $m_{\rm
            d}=0.05$. The line styles refer to $M=3\times
            10^7M_{\odot}$ (solid), $M=2\times 10^7M_{\odot}$ (dotted),
            $M=10^7M_{\odot}$ (dashed). The dot-dashed line refers to
            the case of an $M=3\times 10^7M_{\odot}$ halo, when
            fragmentation at the lowest $\lambda$ is allowed. For the
            two lower halo masses fragmentation will not occur.  }
\label{fig:probbh1}
\end{figure*}

\subsection{The case of atomic hydrogen cooling}

First, we consider the case when molecular hydrogen formation is
inhibited and the gas can consequently cool down to $T_{\rm
gas}\approx 4000$ K through atomic hydrogen cooling only (see Appendix
for details). As discussed above, for a given $T_{\rm gas}$, haloes
with larger $T_{\rm vir}$ produce larger central mass
concentrations. However, we also have to require that the disc does not
fragment, which then limits the analysis to $T_{\rm vir}\lesssim 3
T_{\rm gas}\approx 10^4$ K. The typical redshift of collapse of haloes
with this virial temperature is around $z\approx 10$.

In Fig. \ref{fig:probbh1}, we show the results for the two cases
$Q_{\rm c}=2$ and $Q_{\rm c}=3$, respectively. For both cases, the
haloes are assumed to be at a redshift $z=10$ with mass in the range
of 10$^7\,M_{\odot}$. The lines refer to 3 values of the halo mass,
$M=3\times 10^7M_{\odot}$ (solid line), $M=2\times 10^7M_{\odot}$
(dotted line) and $M=10^7M_{\odot}$ (dashed line). While for the two
lower mass cases $T_{\rm vir}\lesssim 1.8 T_{\rm gas}$ and according
to the results of the previous section, fragmentation is not expected
to take place, for the highest mass case the disc might fragment for
the lowest values of $\lambda$. The dot-dashed line refers to the case
where the possibility of fragmentation is included and in this
instance, we assume that these discs simply do not produce any
central mass concentration. 

It can be seen that, especially for the $Q_{\rm c}=3$ case that a
significant fraction of discs host central mass concentrations. For
dark matter haloes of mass $M=3\times 10^7M_{\odot}$, approximately
12\% of discs have a central concentration of mass larger than
$10^5M_{\odot}$, while the same mass concentration can be achieved by
only 6\% of discs hosted in halos with $M=2\times 10^7M_{\odot}$. The
results obtained assuming $m_{\rm d}=0.1$ for $M=10^7M_{\odot}$ (solid
line) and $M=0.7\times 10^7M_{\odot}$ (dotted line) are shown in
Fig. \ref{fig:probbh2}. In both cases the possibility of fragmentation
is included, but it only affects the lowest spin parameter halos, for
which the probability is small to start with. The two panels refer
again to two different choices of $Q_{\rm c}$. As expected, in this
case, where a larger fraction of the halo mass collapses to the disc,
the fraction of gravitationally unstable discs is larger, resulting in
a larger probability of hosting a large mass concentration at the
center.

\subsection{The case of molecular hydrogen cooling}

\begin{figure*}
\centerline{\epsfig{figure=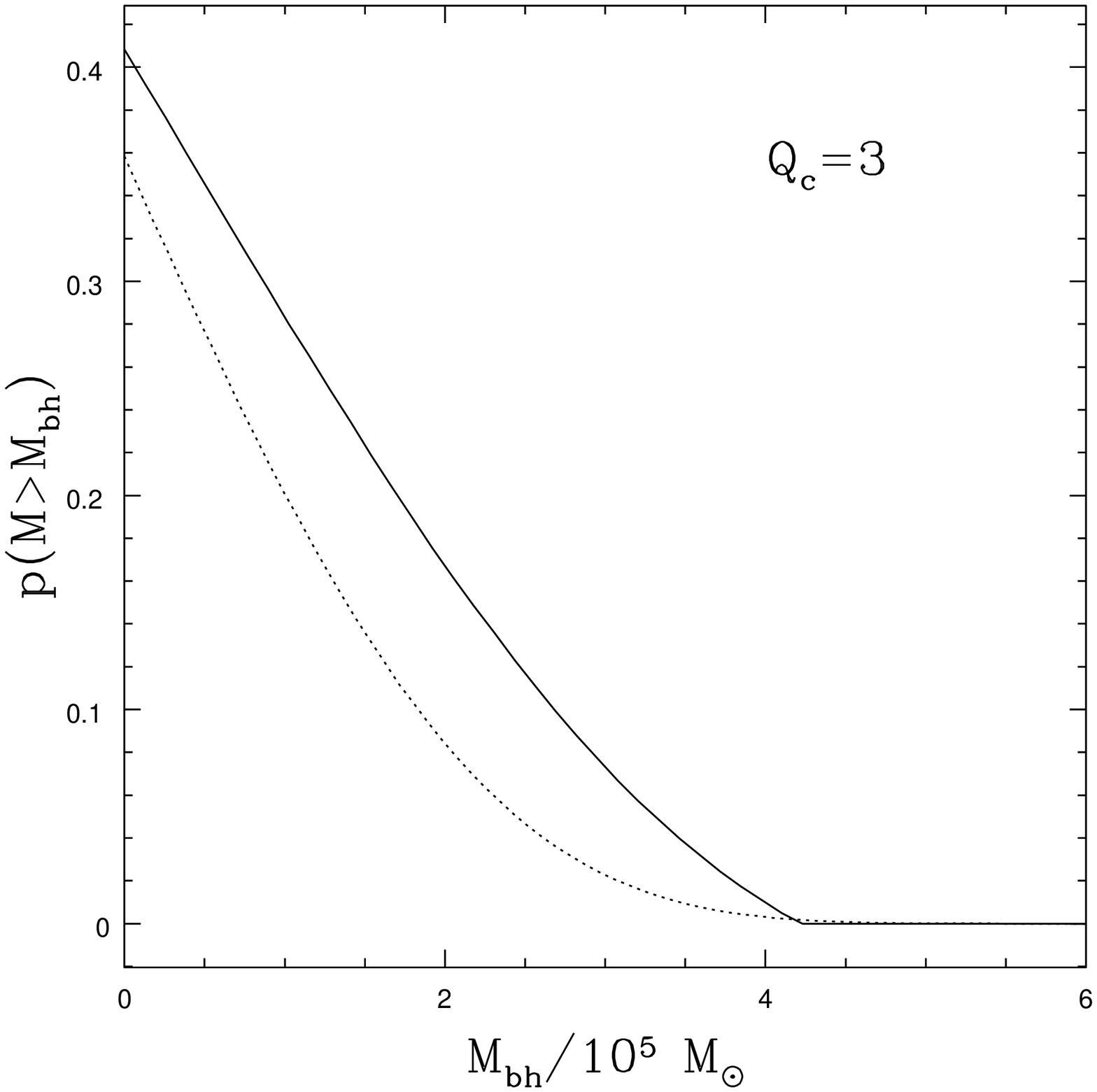,width=0.5\textwidth}
            \epsfig{figure=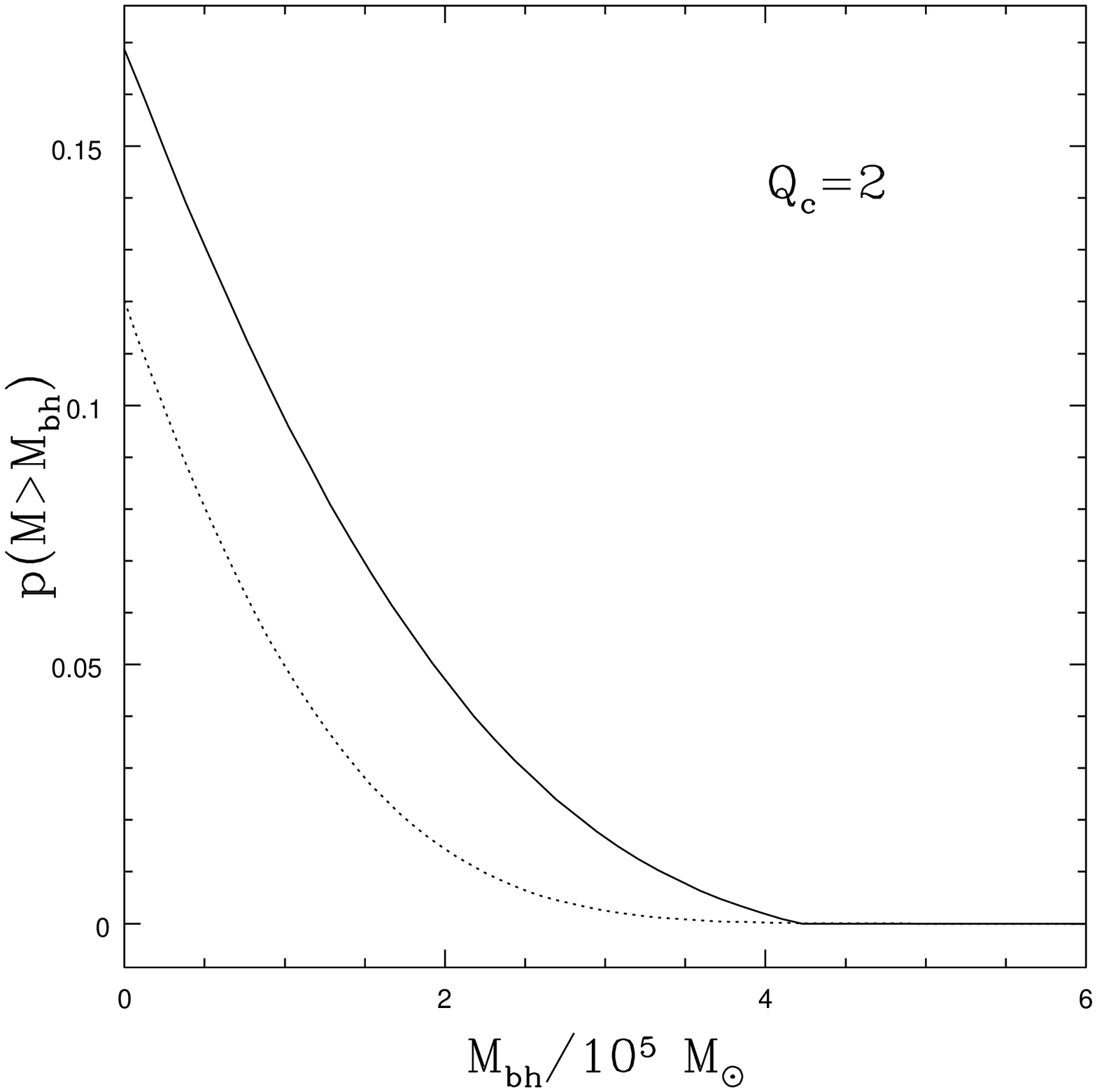,width=0.5\textwidth}}
          \caption{Cumulative distribution of the mass available for
            the formation of a seed black hole. The parent haloes are
            at a redshift $z=10$ and we assume here that $m_{\rm
            d}=0.1$. The line styles refer to haloes with mass
            $M=10^7M_{\odot}$ (solid) and $M=0.6 \times 10^7M_{\odot}$
            (dotted).}
\label{fig:probbh2}
\end{figure*}

We now consider the case where molecular hydrogen can form and the disc
therefore cools down to $T_{\rm gas}\approx 500$ K. In this case, discs
forming out of haloes with $T_{\rm vir}\approx 10^4$ K will most likely
fragment. However, an important point to note is that both the
conditions for fragmentation, and the fraction of halo mass that will
accumulate at the center only depend on the ratio $T_{\rm gas}/T_{\rm
vir}$ as shown in eqn. (\ref{eq:accreted}). Therefore, the very same
central mass accumulation $m_{\rm a}$ that can be obtained for a $4000$
K disc can be obtained for a disc with a temperature ten times lower,
provided that the halo virial temperature is reduced by the same amount
(i.e., from $T_{\rm vir}\approx 10^4$ to $T_{\rm vir}\approx
10^3$). The typical redshift for the collapse of such haloes is $z \sim
20$, higher than the $z\sim 10$ assumed above. Since the halo mass
scales as \mbox{$M\propto T_{\rm vir}^{3/2}(1+z)^{-3/2}$}, the same
results shown in Figs. \ref{fig:probbh1} and \ref{fig:probbh2} would
still hold for a $T_{\rm gas}\approx 500$ K disc, provided that the
mass scale (of both the halo and of the central mass) is reduced by a
factor $\sim 100$. The typical mass of a central concentration in this
case is of the order of $10^3M_{\odot}$. Our proposed mechanism might
therefore also be relevant to the conditions just prior to the
formation of the first stars.

\section{Time-dependent models}
\label{sec:time}

In this section, we develop some time-dependent models to confirm the
results obtained with the simple estimates in previous sections. The
evolution of the surface density $\Sigma$ in a viscous disc with a flat
rotation curve is given by:

\begin{equation}
\frac{\partial\Sigma}{\partial t}=\frac{1}{R}\frac{\partial^2}{\partial
R^2}\left(\nu\Sigma R\right)+\dot{\Sigma}_{\rm in},
\end{equation}
where $\nu$ is the viscosity and $\dot{\Sigma}_{\rm in}$ is a source
term representing the inflow from the halo.

\begin{figure*}
\centerline{\epsfig{figure=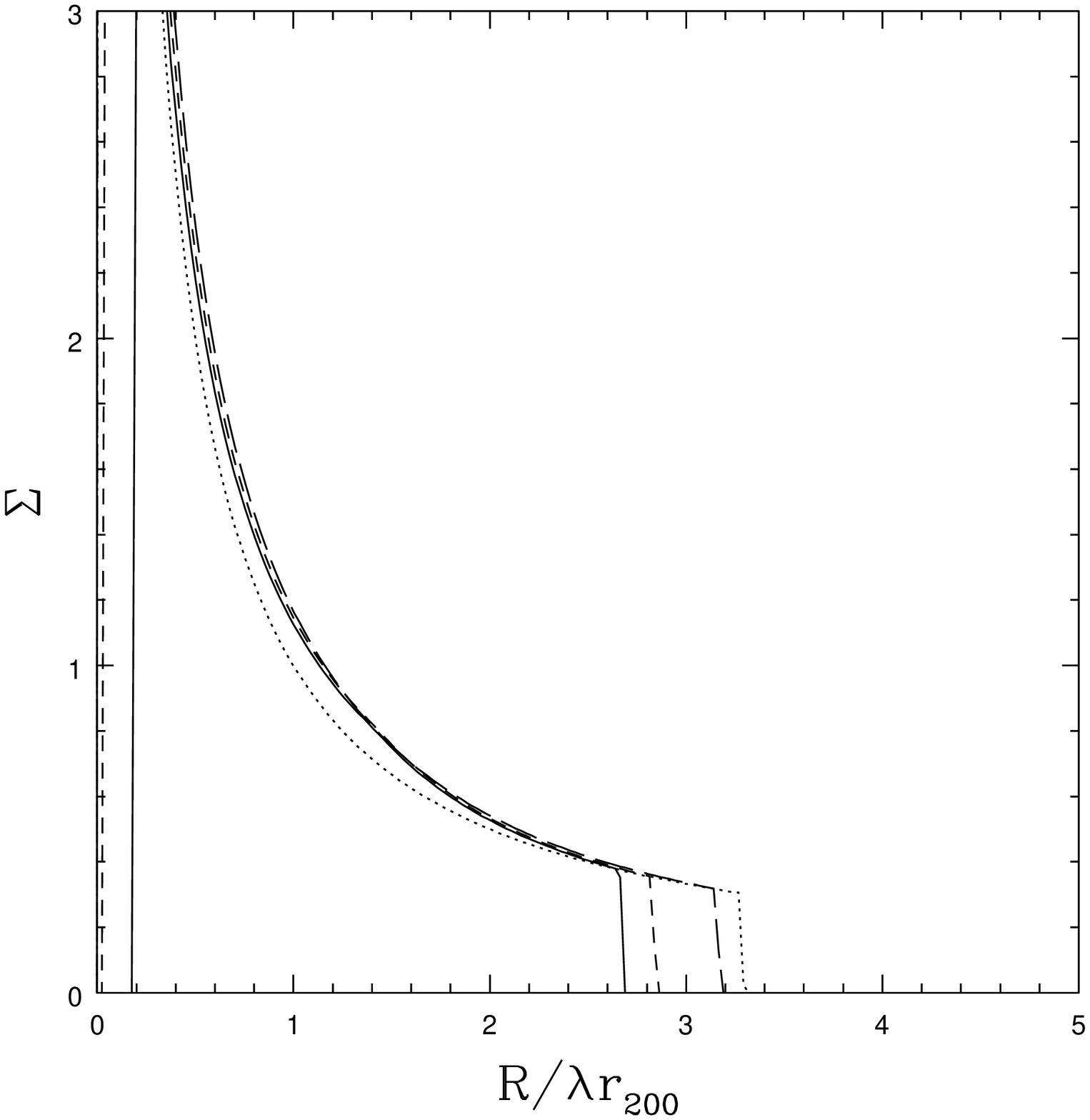,width=0.33\textwidth}
            \epsfig{figure=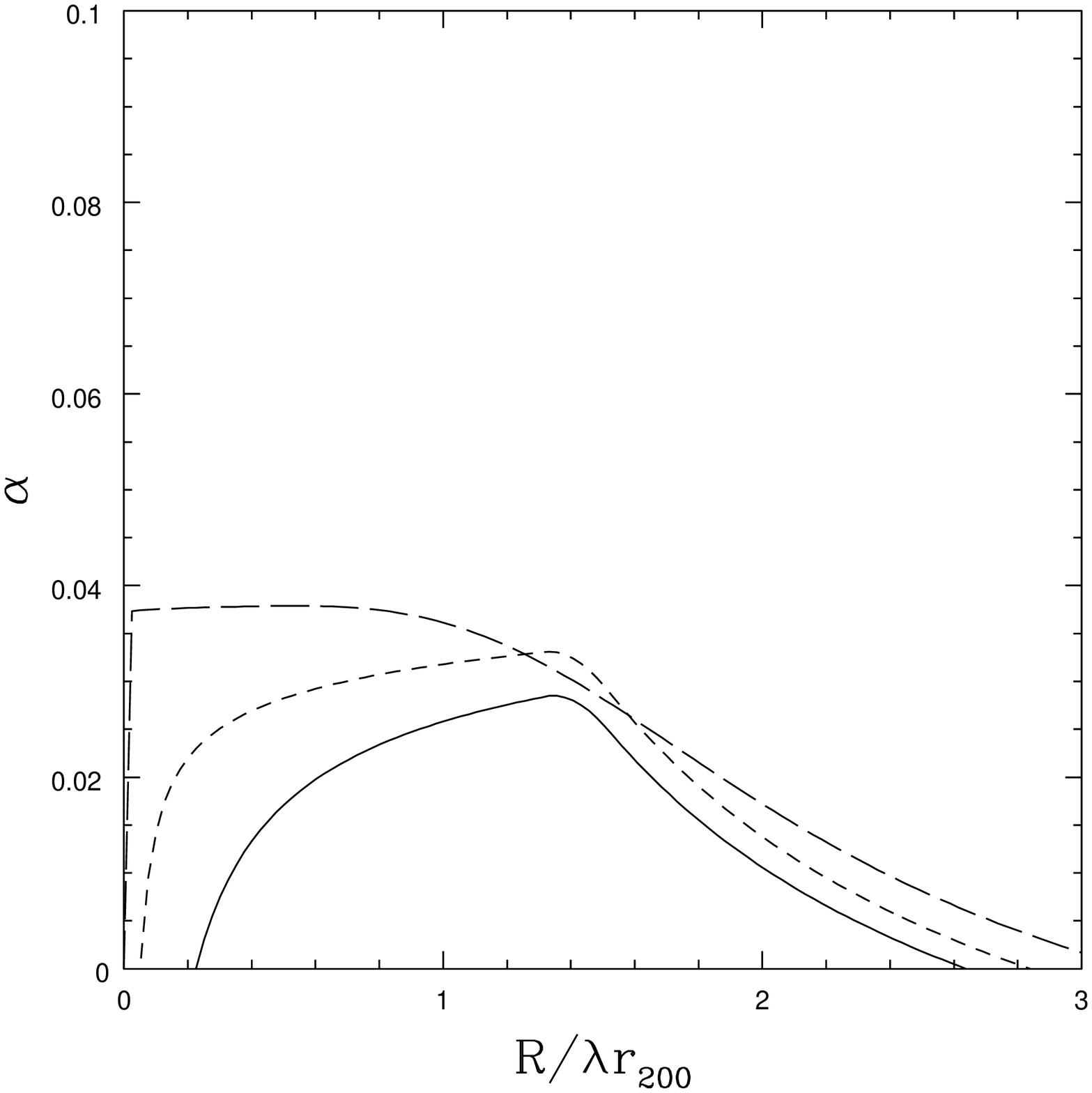,width=0.33\textwidth} 
	    \epsfig{figure=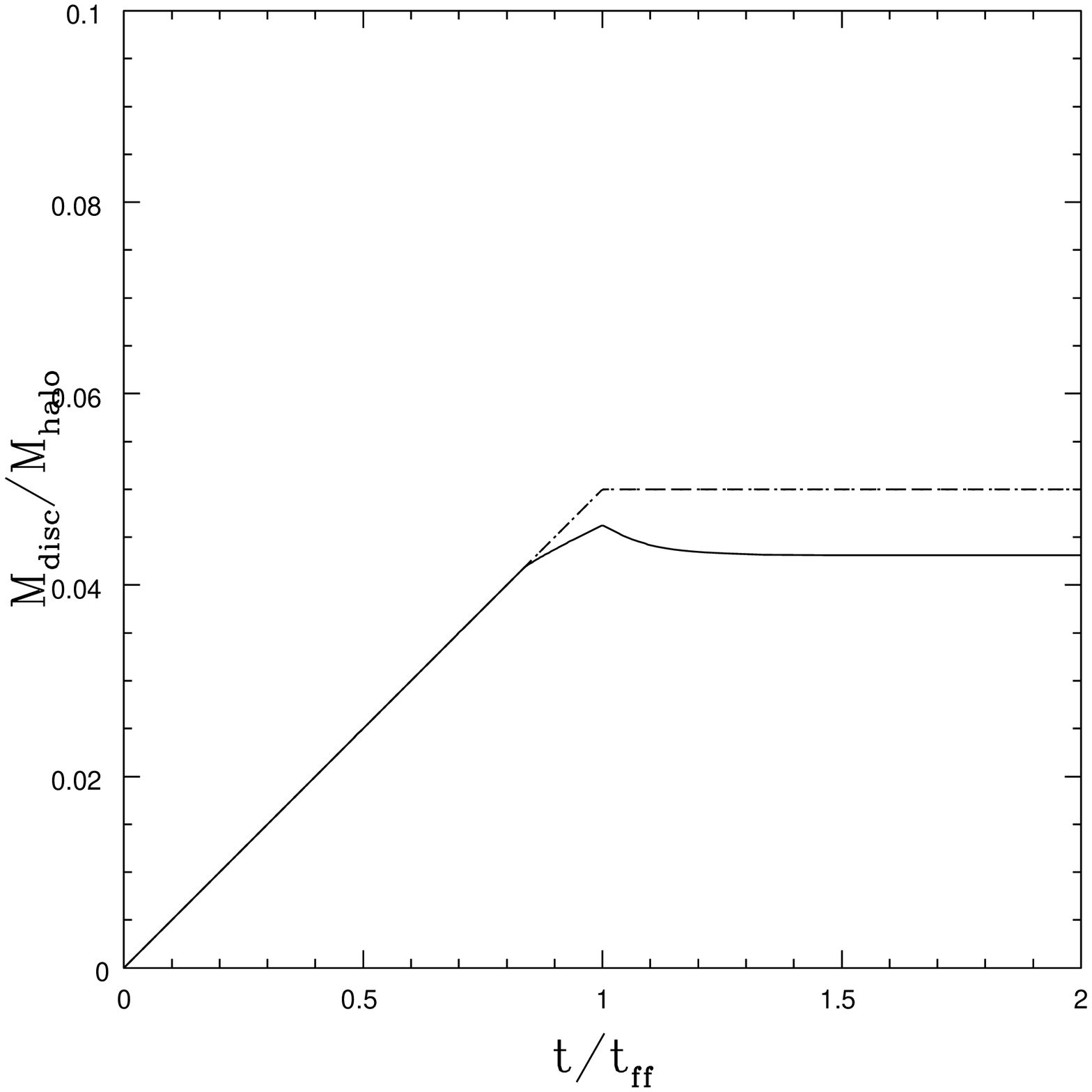,width=0.33\textwidth}}
\caption{Results of a simulation where the mass is added to the disc
with a $\delta$-function, with $\lambda=0.02$, $m_{\rm d}=0.05$,
$c_{\rm s}/V_{\rm h}=0.7$ and $Q_{\rm c}=3$. Left Panel: Surface
density profiles (in units of $160 M_{\odot}/\mbox{pc}^2$, assuming
$T_{\rm gas}=4000$ K and $z=10$) at $t=0.7,0.8,1t_{\rm ff}$ (solid,
short-dashed and long-dashed lines, respectively) and at the end of the
simulation (dotted line). Middle Panel: Profiles of $\alpha_{\rm g}$ at
$t=0.7,0.8,1t_{\rm ff}$ (solid, short-dashed and long-dashed lines,
respectively). Right Panel: Time evolution of the disc mass (solid
line) and of the mass collapsed onto the disc (dot-dashed line).}
\label{fig:time1}
\end{figure*}

\begin{figure*}
\centerline{\epsfig{figure=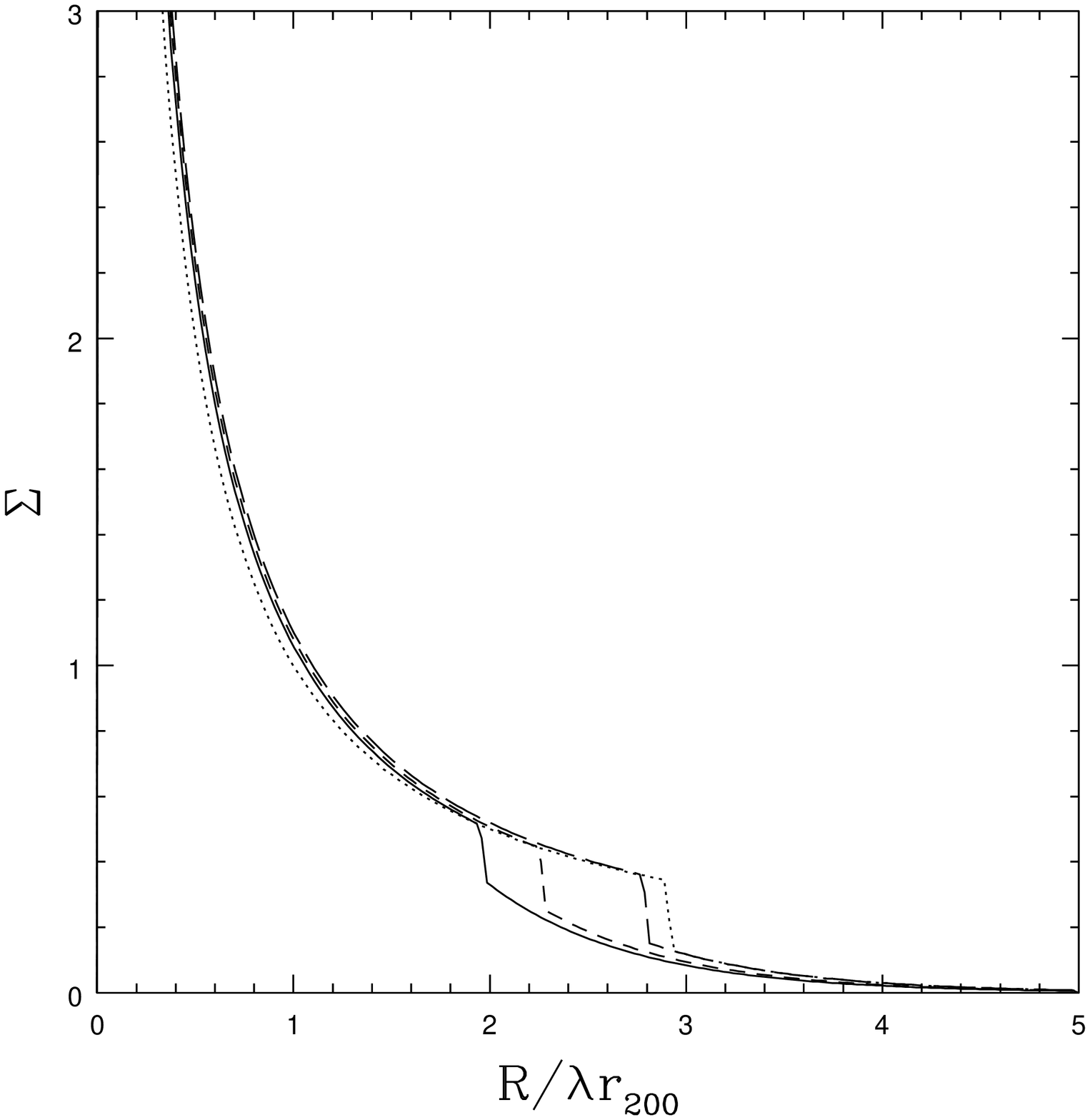,width=0.33\textwidth} 
	    \epsfig{figure=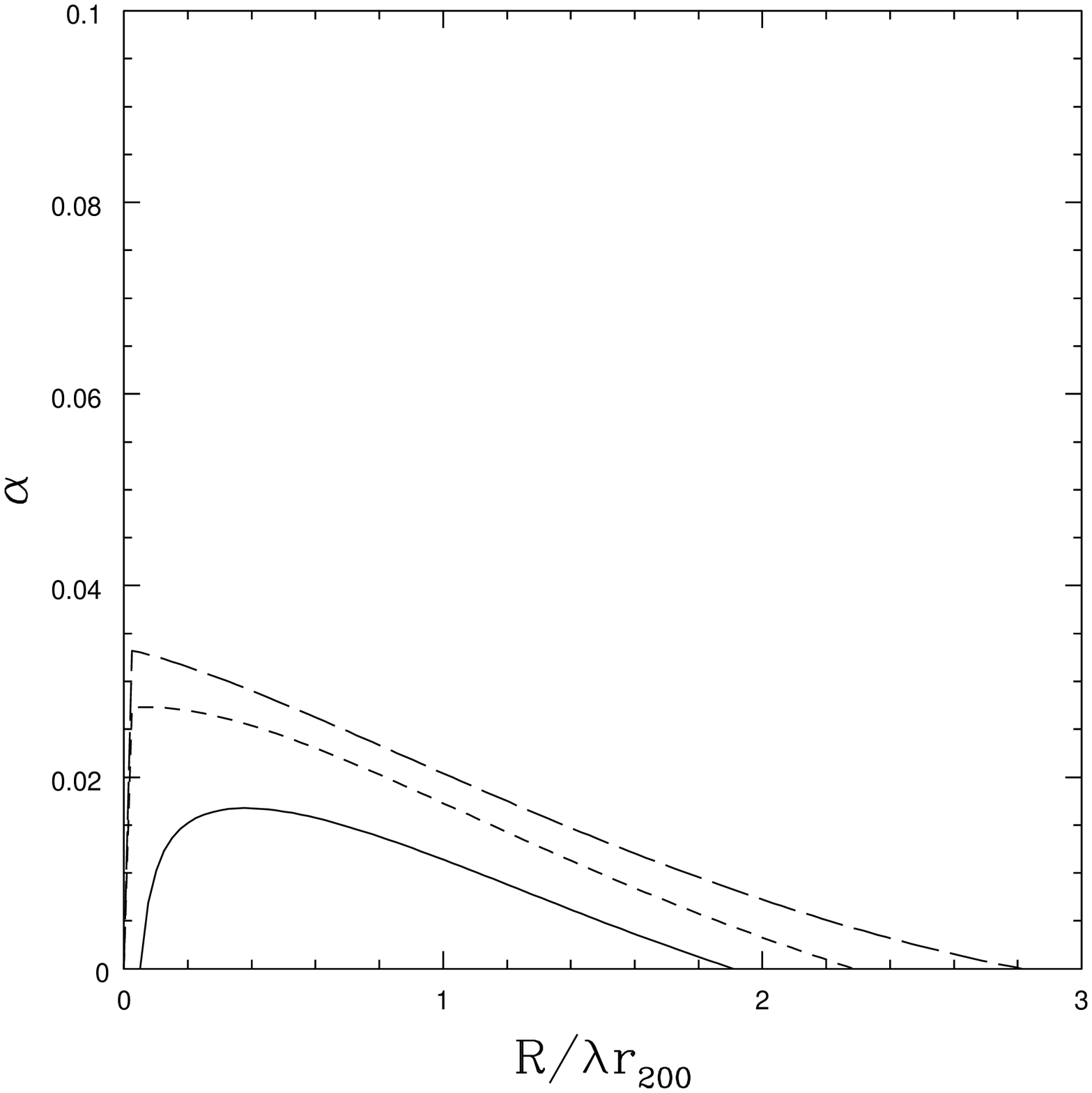,width=0.33\textwidth} 
            \epsfig{figure=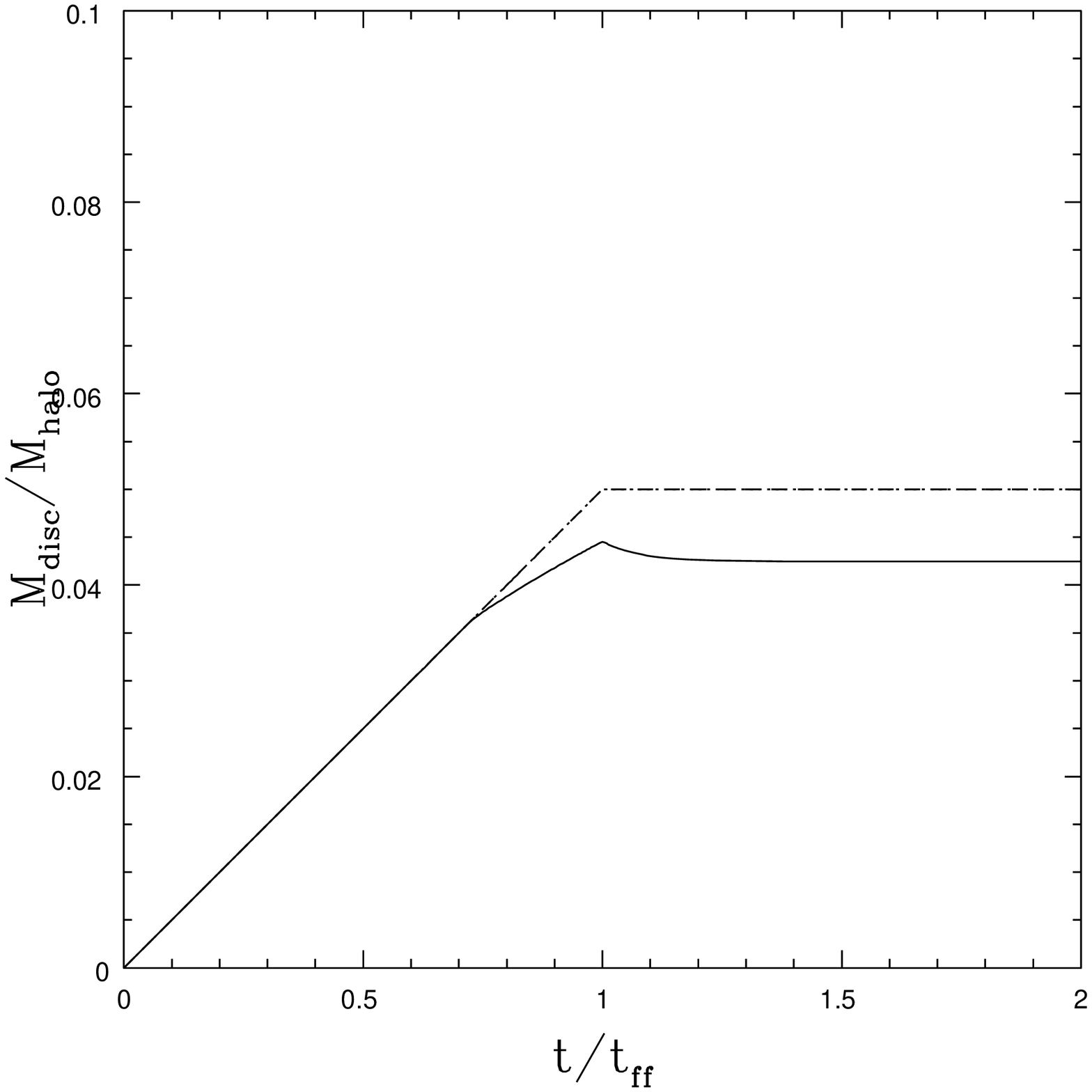,width=0.33\textwidth}}
\caption{Results of a simulation where the mass is added to the disc
with an exponential profile, with $\lambda=0.02$, $m_{\rm d}=0.05$,
$c_{\rm s}/V_{\rm h}=0.7$ and $Q_{\rm c}=3$. Left Panel: Surface
density profiles (in units of $160 M_{\odot}/\mbox{pc}^2$, assuming
$T_{\rm gas}=4000$ K and $z=10$) at $t=0.7,0.8,1t_{\rm ff}$ (solid,
short-dashed and long-dashed lines, respectively) and at the end of the
simulation (dotted line). Middle Panel: Profiles of $\alpha_{\rm g}$ at
$t=0.7,0.8,1t_{\rm ff}$ (solid, short-dashed and long-dashed lines,
respectively). Right Panel: Time evolution of the disc mass (solid
line) and of the mass collapsed onto the disc (dot-dashed line).}
\label{fig:time2}
\end{figure*}

It has been shown \citep{gammie01,LR04} that the transport associated
with the development of such gravitational instabilities can be well
described in terms of an effective viscosity, i.e. within the framework
of the standard $\alpha$ description of accretion discs. It is then
convenient to adopt the simple prescription for viscosity induced by
gravitational instability described by \citet{linpringle90}:
\begin{equation}
\nu=\alpha_{\rm g}c_{\rm s}H=\alpha_{\rm g}\frac{c_{\rm s}^3}{\pi G\Sigma},
\end{equation}
where $\alpha_{\rm g}$, the stress provided by gravitational
instabilities, is modeled as:
\begin{equation}
\alpha_{\rm g}=\eta\left(\frac{Q_{\rm c}^2}{Q^2}-1\right).
\end{equation}
In the previous equation, $Q_{\rm c}$ is a critical value of $Q$, below
which gravitational instabilities are able to provide a stress in the
disc, and $\eta$ is a parameter that essentially determines how far
from the critical value of $Q$ the disc has to be in order to deliver a
given stress. In practice, the general behaviour of such discs is that
they attain a given value of $Q$, such that the stress provided through
$\alpha_{\rm g}$ is the value required to pass on the incoming flux
$\dot{\Sigma}_{\rm in}$. Modifying the value of $\eta$ merely changes
the equilibrium value of $Q$ at which this stress is
provided. Effectively, apart from small changes to the normalization of
$\Sigma$, the specific value of $\eta$ does not affect our results,
therefore confirming that the final state of the disc and the central
accumulation of matter does not depend on the specific viscosity
mechanism. We have generally chosen $\eta=0.1$, but have also explored
other values and do not notice any appreciable difference.

The mass input is modeled in such a way that the accretion rate from
the baryons contained in the halo onto the disc $\dot{M}_{\rm
h}=2\pi\int\dot{\Sigma}_{\rm in}R\mbox{d}R$ is given by
eqn.~(\ref{eq:mhalo}) for $t<t_{\rm ff}=r_{200}/V_{\rm h}$ and vanishes
for $t>t_{\rm ff}$. In this way, the total mass accreted onto the disc
is a fraction $m_{\rm d}$ of the halo mass. The radial dependence of
$\dot{\Sigma}_{\rm in}$ is determined under the constraint that the
total accreted angular momentum is a fraction $j_{\rm d}$ of the halo
angular momentum, with the assumption that $j_{\rm d}=m_{\rm d}$. This
constraint, however, still leaves some freedom in the choice of
$\dot{\Sigma}$. For example, the same result is obtained in the two
extreme cases where all the mass is added to the disc as a
$\delta$-function centered at $R_0=\sqrt{2}\lambda r_{200}$, or the
case where $\dot{\Sigma}_{\rm in}$ has an exponential dependence on
radius $\propto \exp(-R/R_0)$, with $R_0=\lambda r_{200}/\sqrt{2}$.

The results of two of these time-dependent calculations arre shown in
figs. \ref{fig:time1} and \ref{fig:time2} . Both of them refer to the
following choice of parameters: $\lambda=0.02$, $m_{\rm d}=0.05$,
$c_{\rm s}/V_{\rm h}=0.7$, $Q_{\rm c}=3$. Fig. \ref{fig:time1} refers
to the case where the matter is added as a $\delta$-function, while
Fig. \ref{fig:time2} refers to the case where an exponential profile of
$\dot{\Sigma}_{\rm in}$ is adopted. The left panels show the surface
density in the disc at $t=0.7,~0.8, ~1t_{\rm ff}$ and at the end of the
simulation, when all the matter has collapsed to the disc and the disc
becomes stable and is not accreting. The middle panels show the
profiles of $\alpha_{\rm g}$ at $t=0.7,~0.8, ~1t_{\rm ff}$. The right
panels show the time evolution of the disc mass (solid line) and of the
total mass accreted from the halo (dot-dashed line). The fraction of
halo mass accreted in the center at the end of the simulation was
$m_{\rm a}\approx 0.00688$ for the $\delta$-function case and $m_{\rm
a}\approx 0.0073$ for the exponential case, in very good agreement with
the simple estimates obtained from eqn.~(\ref{eq:accreted}).

\begin{figure}
\centerline{\epsfig{figure=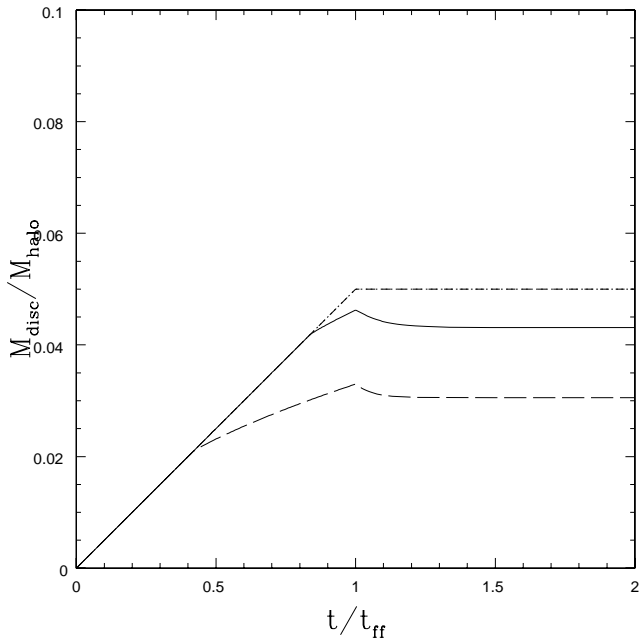,width=0.5\textwidth}}            
\caption{Comparison of the time evolution of the disc mass for two
simulations with different values of the halo spin parameter: 
$\lambda=0.02$ (solid line) and $\lambda=0.01$ (dashed line). The dot-dashed 
line shows as in Figs. \ref{fig:time1} and \ref{fig:time2} the total mass 
collapsed onto the disc.}
\label{fig:time3}
\end{figure}

A comparison of the evolution of disc mass for two cases with different
values of $\lambda=0.02$ (black line) and $\lambda=0.01$ (dashed line)
in shown in fig. \ref{fig:time3}. The other parameters were set as
before and the mass input function was a $\delta$-function. The
accreted mass at the end of the simulation with $\lambda=0.01$ is
$m_{\rm a}=0.019$, once again in very good agreement with
eqn.~(\ref{eq:accreted}).

An interesting feature that appears from these time-dependent models,
is the value of the viscosity parameter $\alpha_{\rm g}$ during the
simulation. If the disc were to deliver a steady mass accretion rate of
$\dot{M}_{\rm h}$, as given by eqn.~(\ref{eq:mhalo}), it would require
an $\alpha_{\rm g}=m_{\rm d}/2(V_{\rm h}/c_{\rm s})^3\approx 0.07$ for
the chosen parameters (comparing eqn.~(\ref{eq:mhalo}) with
eqn.~(\ref{eq:mdot})). However, since the system is not in a steady
state, the actual value of $\alpha_{\rm g}$ needed to provide the
required mass flux is actually lower, peaking at roughly $\alpha_{\rm
g}\approx 0.04$. This is quite important to determine whether we expect
the disc to fragment and form stars or not. As described earlier, we
expect fragmentation whenever the stress needed to transport the mass
flux through the disc is larger than a threshold value $\alpha_{\rm
c}\approx 0.06$. With the parameters chosen above, that requires
$\alpha\approx 0.07$, we would then expect fragmentation, whereas in
fact an evolving disc will always have a lower value effective of
$\alpha$ and is therefore not expected to fragment.

The above behaviour can be understood quantitatively as follows: if
some matter $\delta M$ is delivered at a radius $R_0$, only a fraction
$f=(1-R_0/R_{\rm d})$ will flow into the inner disc, while the
remaining $\delta M (R_0/R_{\rm d})$ will be carried out to compensate
for the angular momentum lost by infalling matter. Of course, this will
leave an unbalanced mass flux in the outer disc that will lead to an
increase of $R_{\rm d}$. This however happens on a very long
time-scale, especially since the viscosity in the outer disc becomes
very small. Using eqn.~(\ref{eq:rd}) to obtain the outer disc radius
and assuming $R_0=\sqrt{2}\lambda r_{200}$, we obtain:
\begin{equation}
f=\left(1-\frac{\sqrt{2}\lambda r_{200}}{R_{\rm d}}\right)=
\frac{1}{2}\left(1+\frac{m_{\rm a}}{m_{\rm d}}\right), 
\end{equation}
where we have also assumed $j_{\rm d}=m_{\rm d}$. For the case
considered in Fig. \ref{fig:time1}, we obtain $f\approx 0.57$ which
then lowers the required stress from $\alpha\approx0.07$ to
$\alpha\approx0.04$, in perfect agreement with the results of the
simulation. We have therefore demonstrated the result derived in
Section \ref{sec:frag} that the mass flow into the inner disc during
the disc build up is actually smaller than its steady-state value by a
factor $f$.

\section{Luminosity and observability} 

In this section we estimate the luminosity and potential observability
these pre-galactic discs. Here we only concentrate on the hot discs
dominated by atomic hydrogen cooling. The dissipation rate per unit
surface area determined by the accretion process is:

\begin{equation}
D(R)=\nu\Sigma(R\Omega')^2=\frac{\mdot}{2\pi}\left(\frac{V_{\rm h}}{R}
\right)^2, 
\end{equation}
where we have assumed a flat rotation curve at $V_{\rm h}$. The total
luminosity emitted by the disc is:
\begin{eqnarray}
\nonumber L_{\rm disc} & = & 2\pi\int_{R_{\rm in}}^{R_{\rm out}}RD(R)\de R \\
& =  &\mdot
V_{\rm h}^2 \int_{R_{\rm in}}^{R_{\rm out}}\frac{\de R}{R} \approx m_{\rm d} 
\frac{V_{\rm h}^5}{G} \ln(R_{\rm out}/R_{\rm in}),
\label{eq:luminosity}
\end{eqnarray}
where $R_{\rm in}$ and $R_{\rm out}$ are the inner and outer disc
radius, respectively, using $\mdot\approx m_{\rm d}
V_{\rm h}^3/G$. The outer radius of the disc is given by $R_{\rm
  out}=R_{\rm d}\approx 100$ pc, for a gravitationally unstable disc,
with $\lambda=0.05$ at $z=10$. The inner radius is harder to
estimate. Our model will not be valid in the innermost
parts of the disc, where the influence of a growing black hole will
become important. Inside this radius the rotation curve will start to
rise in an approximately Keplerian way, and the disc will therefore
become hotter. This inner limiting radius can be estimated from:
\begin{equation}
R_{\rm in}=\frac{GM_{\rm BH}}{V_{\rm h}^2}\approx 4.5\times 10^{-2}
\left(\frac{M_{\rm BH}}{10^3M_{\odot}}\right)
\left(\frac{10\mbox{km/sec}}{V_{\rm h}}\right)^2 \mbox{pc}.
\end{equation}
Substituting these values into eqn.~(\ref{eq:luminosity}), we find that the
luminosity arising from the outer part of the disc is:
\begin{equation}
\label{eq:lumi}
L_{\rm disc}\approx 4000 L_{\odot} \left(\frac{V_{\rm
h}}{10\mbox{km/sec}}\right)^5.
\end{equation}

Note that, while in principle eqn. (\ref{eq:lumi}) would tell us that
larger values of $V_{\rm h}$ would provide a much larger luminosity,
this is actually unlikely, since for larger $V_{\rm h}$ the disc would
efficiently fragment into stars rather than accrete at a higher rate,
as discussed above. We therefore argue that the accretion luminosity in
these discs is unlikely to exceed the estimate above.  Since the gas
cools mainly through atomic hydrogen line emission, most of this
luminosity will be emitted in the Lyman series, and in particular in
Ly$\alpha$.

\begin{figure}
\centerline{\epsfig{figure=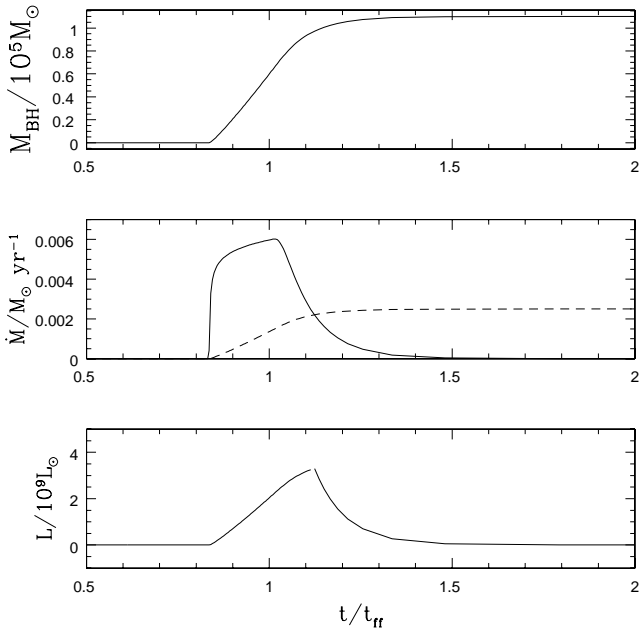,width=0.5\textwidth}}            
\caption{Upper Panel: mass of the central concentration as a function
of time for the case shown in Fig. \ref{fig:time1}. Middle Panel:
$\dot{M}$ in the inner disc (solid line) and the Eddington rate for a
black hole of mass $M_{\rm BH}$ (dotted line). Bottom Panel: expected
luminosity as a function of time, assuming that it is limited by the
Eddington value at early times and by mass supply at late times.}
\label{fig:edd}
\end{figure}

In fact, most of the luminosity of {\it the system} will be released in
the inner disc, as the matter falls to the bottom of the potential well
to ultimately form and feed the growing black hole. A detailed
description of this process is beyond the scope of the present paper
and has been discussed elsewhere
\citep{volonteri05b,begelman06}. Indeed, the accretion rate onto the
black hole or onto the `quasi-star' described in such models is one of
their input parameters.  This parameter can be computed
self-consistently in a time-dependent way based on our calculation
presented in section \ref{sec:time}. The results are shown in
Fig. \ref{fig:time1} and Fig. \ref{fig:edd}. The upper panel of
Fig. \ref{fig:edd} shows the time evolution of the mass of the growing
central object $M_{\rm BH}$. The middle panel shows the evolution of
the mass accretion rate onto it (solid line) and the corresponding
Eddington rate for a black hole of mass $M_{\rm BH}$ with an efficiency
of matter-luminosity conversion set equal to 0.1. As can be seen the
initial phases are characterized by a strongly super-Eddington
accretion phase (as expected, and as predicted by
\citealt{volonteri05b,begelman06}). At later times, however, as the
mass of the central object grows to reach roughly its asymptotic value,
the accretion rate decreases, becoming sub-Eddington when the black
hole has a mass of roughly $8\times 10^4M_{\odot}$ in this case. The
bottom plot shows the luminosity expected to be radiated from such a
system, assuming that during the early super-Eddington phase the output
luminosity is still limited by the Eddington value, while in the
subsequent phase it is limited by the decrease in the mass supply. The
peak of the luminosity is expected to occur just before the transition
from the super- to the sub-Eddington regime and it corresponds to a few
times $10^9 L_{\odot}$.

\subsection{Prospects for observing these pre-galactic discs}

Given the low intrinsic luminosity of these pre-galactic discs ($\sim
4000\,L_{\odot}$), the prospects for direct detection are not
promising.  However, including the accretion luminosity of the black
holes that are being assembled in the centers of these discs, they are
likely to be detectable by the James Webb Space Telescope (JWST). The
best strategy for detecting these objects is to exploit their lensing by
foreground clusters. Using massive, foreground cluster-lenses as
gravitational telescopes, we can expect to detect highly magnified
pre-galactic discs. Therefore, a survey through cluster lenses by JWST is
likely to be the most viable detection strategy.
  
\section{Discussion and conclusions}

In this paper, we demonstrate that massive central concentrations can
form naturally from pre-galactic discs that assemble in dark matter
halos at high redshift. The masses of these concentrations depend on
key properties of the host halo, mass, spin and gas cooling in the
halo. In particular, low spin halos and massive halos are most
efficient in concentrating gas in their centers. However, not all halos
will be able to accrete gas into their centers. Taking into account the
possibility of fragmentation furthers restricts the formation of
eventual seed black holes.  Using simple stability criteria, we predict
that fragmentation and subsequent star formation are the fate for gas
in some fraction of haloes. We derive 3 interesting regimes that are
determined by the ratio of $T_{\rm vir}/T_{\rm gas}$, (i) if this ratio
is greater than 2.9, haloes will form discs that will fragment and form
stars and not directly form black holes; (ii) if this ratio lies
between 1.8 - 2.9, haloes with low spin will lead to central mass
concentrations; however, the lowest spin cases might be affected by
fragmentation; (iii) if this ratio is lower than 1.8 the haloes will
not produce fragmenting discs but will successfully accrete gas in
their centers.  Calculating the accretion rates for the fate of the
accumulated gas we find super-Eddington rates leading to eventual black
holes of masses up to $10^5 M_{\odot}$.

The model proposed here has two important predecessors in the work of
\citet{eisenstein95} and in that of \citet{koushiappas04}, but also
presents significant differences with respect to both of them.
Primordial gas acquires angular momentum through tidal interaction with
the surrounding \citep{peebles69}. This is generally measured through
the parameter $\lambda$, the distribution of which can be determined
from numerical simulations \citep{warren92}. Thus, the centrifugal
barrier is the main obstacle to the formation of any compact object in
the center of primordial galaxies. It is therefore not surprising that
all models (including our own) predict that the most favourable sites
for black hole formation are haloes with low spin. It is also not
surprising that any black hole formation model has to deal with the
problem of angular momentum removal. This issue was tackled in the work
of \citet{eisenstein95}, who calculated the viscous time-scale for
primordial thin discs and assumed that black hole formation would only
occur when the discs are so compact that their viscous time-scale is
less than the typical time-scale for star formation. They found that
only very rare haloes, with $\lambda$ smaller by at least a factor 30
with respect to the average spin, would be able to produce black
holes. This is substantially different from what we find here. Indeed,
we find that even haloes with $\lambda$ as large as 0.02-0.03 do
produce large central mass concentrations. There are two main aspects
with respect to which our model differs substantially from
\citet{eisenstein95}. The first one is the nature of the viscosity
mechanism. \citet{eisenstein95} explicitly neglect the contribution of
gravitational instabilities and assume that viscosity is driven by
``turbulence''. In standard accretion discs, turbulence is thought to
arise from MHD instabilities \citep{balbusreview}. However, for such
primordial discs, where the primordial magnetic field is very weak and
the gas is predominantly atomic or molecular, it is unlikely that MHD
instabilities would be dynamically important. The second, and most
important, difference is that \citet{eisenstein95} considered
relatively thin discs, with $H/R\approx 0.03$, whereas our discs are
substantially thicker. Since the viscous timescale is proportional to
$(H/R)^{-2}$, this means that the viscous timescale as estimated by
\citet{eisenstein95} is much larger than in our case. This in turn, is
what leads to their more pessimistic estimate of the rarity of black
hole forming haloes. Finally, a significant difference between our
model and that of \citet{eisenstein95} is that our model naturally
leads to a robust determination of the seed black hole mass
function.

\citet{koushiappas04} propose that black hole seeds with masses of the
order of $10^5M_{\odot}$ (i.e. very similar to those obtained here) can
form out of low angular momentum material in massive haloes. They
assume that the lowest angular momentum material within the halo forms
a compact disc which is gravitationally unstable and accretes onto the
center due to the effect of an effective viscosity driven by the
instability, in a way similar to what we propose here. However, they do
not consider self-consistently the evolution of the surface density
profile induced by the assumed viscosity mechanism. In their picture,
the disc surface density simply grows linearly with time (at the same
rate, independent of the radius) and at any given time it reflects the
original angular momentum distribution of the gaseous component of the
halo. On the contrary, what determines the surface density profile is
actually the viscosity mechanism (which, being related to gravitational
instabilities, in turn only depends on $Q$). As we have shown in
Section \ref{sec:time} (Figs. \ref{fig:time1} and \ref{fig:time2}),
where we numerically follow the evolution of such discs, the final
$\Sigma$ profile (and the total accreted mass $m_{\rm a}$) is the same
whether we add mass to the disc with an exponential profile or with a
$\delta$-function, representing two extreme cases in the original
angular momentum distribution within the halo. As a consequence of
this, the estimates of black hole masses and their distribution given
by \citet{koushiappas04} artificially depend on the initial angular
momentum distribution and also on the assumed viscosity law. On the
other hand, we have also clearly shown that, as long as viscosity is
driven by gravitational instabilities, the black hole mass distribution
is independent of viscosity.

In the present paper, for illustrative purposes, we have taken the
simplifying assumption that these primordial discs live in the
potential well of a simple isothermal sphere, with a given circular
velocity $V_{\rm h}$. For a more realistic density profile, the NFW
profile \citep{nfw}, in its innermost parts, the disc mass might
dominate the potential well and become radially unstable, giving rise
to bar-like instabilities. This, however, will leave our results and
conclusions unaffected. Firstly, the instability criterion for such
global instabilities \citep{christo95} is rather similar to our adopted
criterion based on $Q$, and marginal stability is found to occur for
equivalent values of $Q_{\rm c}\approx 2-3$, as adopted here. Secondly,
since the disc mass is only a small fraction of the halo mass, only the
innermost parts of the disc will be subject to bar-like instabilities,
at radii much smaller than the typical disc radius $R_{\rm d}$. The
development of such instabilities might in fact enhance the accretion
rate in the inner disc, but will not change the estimates of the total
accreted mass obtained here.  The important and interesting consequence
of our model is that black hole masses of $10^9 M_{\odot}$ powering the
luminous SDSS quasars at $z = 6$ can form comfortably within the
available time of 1 Gyr in the concordance cosmology from our seed
masses of $10^5 M_{\odot}$ at $z \sim 10$.

\section*{Acknowledgements}

The authors acknowledge useful conversations and comments from Philip
Armitage, Mitch Begelman, Cathie Clarke, Micheal Mayer, Jim Pringle and 
Marta Volonteri.

\bibliographystyle{mn2e} 
\bibliography{lodato}

\begin{thebibliography}{}

\bibitem[\protect\citeauthoryear{{Abel}, {Bryan} \& {Norman}}{{Abel}
  et~al.}{2000}]{abel00}
{Abel} T.,  {Bryan} G.~L.,    {Norman} M.~L.,  2000, ApJ, 540, 39

\bibitem[\protect\citeauthoryear{Balbus \& Hawley}{Balbus \&
  Hawley}{1998}]{balbusreview}
Balbus S.~A.,  Hawley J.~F.,  1998, Reviews of Modern Physics, 70, 1

\bibitem[\protect\citeauthoryear{{Begelman}, {Volonteri} \& {Rees}}{{Begelman}
  et~al.}{2006}]{begelman06}
{Begelman} M.~C.,  {Volonteri} M.,    {Rees} M.~J.,  2006, ArXiv Astrophysics
  e-prints

\bibitem[\protect\citeauthoryear{Bertin \& Lodato}{Bertin \&
  Lodato}{1999}]{BL99}
Bertin G.,  Lodato G.,  1999, A\&A, 350, 694

\bibitem[\protect\citeauthoryear{{Bromm}, {Coppi} \& {Larson}}{{Bromm}
  et~al.}{2002}]{bromm02}
{Bromm} V.,  {Coppi} P.~S.,    {Larson} R.~B.,  2002, ApJ, 564, 23

\bibitem[\protect\citeauthoryear{{Bromm} \& {Loeb}}{{Bromm} \&
  {Loeb}}{2003}]{bromm03}
{Bromm} V.,  {Loeb} A.,  2003, ApJ, 596, 34

\bibitem[\protect\citeauthoryear{{Cen}}{{Cen}}{1992}]{cen1992}
{Cen} R.,  1992, ApJS, 78, 341

\bibitem[\protect\citeauthoryear{{Christodoulou}, {Shlosman} \&
  {Tohline}}{{Christodoulou} et~al.}{1995}]{christo95}
{Christodoulou} D.~M.,  {Shlosman} I.,    {Tohline} J.~E.,  1995, ApJ, 443, 563

\bibitem[\protect\citeauthoryear{{Dalcanton}, {Spergel} \&
  {Summers}}{{Dalcanton} et~al.}{1997}]{dalcanton97}
{Dalcanton} J.~J.,  {Spergel} D.~N.,    {Summers} F.~J.,  1997, ApJ, 482, 659

\bibitem[\protect\citeauthoryear{{Di Matteo}, {Springel} \& {Hernquist}}{{Di
  Matteo} et~al.}{2005}]{dimatteo05}
{Di Matteo} T.,  {Springel} V.,    {Hernquist} L.,  2005, Nature, 433, 604

\bibitem[\protect\citeauthoryear{{Eisenstein} \& {Loeb}}{{Eisenstein} \&
  {Loeb}}{1995}]{eisenstein95}
{Eisenstein} D.~J.,  {Loeb} A.,  1995, ApJ, 443, 11

\bibitem[\protect\citeauthoryear{{Fall} \& {Efstathiou}}{{Fall} \&
  {Efstathiou}}{1980}]{fall80}
{Fall} S.~M.,  {Efstathiou} G.,  1980, MNRAS, 193, 189

\bibitem[\protect\citeauthoryear{Fan, Hennawi, Richards, Strauss, Schneider,
  Donley, Young, Annis, Lin, Lampeitl, Lupton, Gunn, Knapp, Brandt, Anderson,
  Bahcall, Brinkmann, Brunner, Fukugita, Szalay, Szokoly \& York}{Fan
  et~al.}{2004}]{fan04}
Fan X.,  Hennawi J.,  Richards G.,  Strauss M.,  Schneider D.,  Donley J.,
  Young J.,  Annis J.,  Lin H.,  Lampeitl H.,  Lupton R.~H.,  Gunn J.,  Knapp
  G.,  Brandt W.,  Anderson S.,  Bahcall N.,  Brinkmann J.,  Brunner R.,
  Fukugita M.,  Szalay A.,  Szokoly G.,    York D.,  2004, AJ, 128, 515

\bibitem[\protect\citeauthoryear{Ferrarese \& Merritt}{Ferrarese \&
  Merritt}{2000}]{ferrarese2000}
Ferrarese L.,  Merritt D.,  2000, ApJ, 539, 9

\bibitem[\protect\citeauthoryear{{Galli} \& {Palla}}{{Galli} \&
  {Palla}}{1998}]{galli98}
{Galli} D.,  {Palla} F.,  1998, A\&A, 335, 403

\bibitem[\protect\citeauthoryear{Gammie}{Gammie}{2001}]{gammie01}
Gammie C.~F.,  2001, ApJ, 553, 174

\bibitem[\protect\citeauthoryear{{Haehnelt} \& {Rees}}{{Haehnelt} \&
  {Rees}}{1993}]{haehnelt93}
{Haehnelt} M.~G.,  {Rees} M.~J.,  1993, MNRAS, 263, 168

\bibitem[\protect\citeauthoryear{{Katz} \& {Gunn}}{{Katz} \&
  {Gunn}}{1991}]{katz}
{Katz} N.,  {Gunn} J.~E.,  1991, ApJ, 377, 365

\bibitem[\protect\citeauthoryear{{Kauffmann} \& {Haehnelt}}{{Kauffmann} \&
  {Haehnelt}}{2000}]{kauffmann00}
{Kauffmann} G.,  {Haehnelt} M.,  2000, MNRAS, 311, 576

\bibitem[\protect\citeauthoryear{{Koushiappas}, {Bullock} \&
  {Dekel}}{{Koushiappas} et~al.}{2004}]{koushiappas04}
{Koushiappas} S.~M.,  {Bullock} J.~S.,    {Dekel} A.,  2004, MNRAS, 354, 292

\bibitem[\protect\citeauthoryear{Lin \& Pringle}{Lin \&
  Pringle}{1990}]{linpringle90}
Lin D. N.~C.,  Pringle J.~E.,  1990, ApJ, 358, 515

\bibitem[\protect\citeauthoryear{Lodato \& Rice}{Lodato \& Rice}{2004}]{LR04}
Lodato G.,  Rice W. K.~M.,  2004, MNRAS, 351, 630

\bibitem[\protect\citeauthoryear{Lodato \& Rice}{Lodato \& Rice}{2005}]{LR05}
Lodato G.,  Rice W. K.~M.,  2005, MNRAS, 358, 1489

\bibitem[\protect\citeauthoryear{{Loeb} \& {Rasio}}{{Loeb} \&
  {Rasio}}{1994}]{loeb94}
{Loeb} A.,  {Rasio} F.~A.,  1994, ApJ, 432, 52

\bibitem[\protect\citeauthoryear{{Madau} \& {Rees}}{{Madau} \&
  {Rees}}{2001}]{madaurees01}
{Madau} P.,  {Rees} M.~J.,  2001, ApJ, 551, L27

\bibitem[\protect\citeauthoryear{{Magorrian}, {Tremaine}, {Richstone},
  {Bender}, {Bower}, {Dressler}, {Faber}, {Gebhardt}, {Green}, {Grillmair},
  {Kormendy} \& {Lauer}}{{Magorrian} et~al.}{1998}]{magorrian98}
{Magorrian} J.,  {Tremaine} S.,  {Richstone} D.,  {Bender} R.,  {Bower} G.,
  {Dressler} A.,  {Faber} S.~M.,  {Gebhardt} K.,  {Green} R.,  {Grillmair} C.,
  {Kormendy} J.,    {Lauer} T.,  1998, AJ, 115, 2285

\bibitem[\protect\citeauthoryear{{Mapelli}, {Ferrara} \& {Rea}}{{Mapelli}
  et~al.}{2006}]{mapelli06}
{Mapelli} M.,  {Ferrara} A.,    {Rea} N.,  2006, MNRAS, 368, 1340

\bibitem[\protect\citeauthoryear{{Mayer} \& {Duschl}}{{Mayer} \&
  {Duschl}}{2005}]{mayer05}
{Mayer} M.,  {Duschl} W.~J.,  2005, MNRAS, 358, 614

\bibitem[\protect\citeauthoryear{{Mestel}}{{Mestel}}{1963}]{mestel63}
{Mestel} L.,  1963, MNRAS, 126, 553

\bibitem[\protect\citeauthoryear{{Mo}, {Mao} \& {White}}{{Mo}
  et~al.}{1998}]{mo98}
{Mo} H.~J.,  {Mao} S.,    {White} S.~D.~M.,  1998, MNRAS, 295, 319

\bibitem[\protect\citeauthoryear{{Navarro}, {Frenk} \& {White}}{{Navarro}
  et~al.}{1997}]{nfw}
{Navarro} J.~F.,  {Frenk} C.~S.,    {White} S.~D.~M.,  1997, ApJ, 490, 493

\bibitem[\protect\citeauthoryear{{Oh} \& {Haiman}}{{Oh} \& {Haiman}}{2002}]{oh}
{Oh} S.~P.,  {Haiman} Z.,  2002, ApJ, 569, 558

\bibitem[\protect\citeauthoryear{{Peebles}}{{Peebles}}{1969}]{peebles69}
{Peebles} P.~J.~E.,  1969, ApJ, 155, 393

\bibitem[\protect\citeauthoryear{Pringle}{Pringle}{1981}]{pringle81}
Pringle J.~E.,  1981, ARA\&A, 19, 137

\bibitem[\protect\citeauthoryear{{Rice}, {Lodato} \& {Armitage}}{{Rice}
  et~al.}{2005}]{RLA05}
{Rice} W.~K.~M.,  {Lodato} G.,    {Armitage} P.~J.,  2005, MNRAS, 364, L56

\bibitem[\protect\citeauthoryear{{Ricotti} \& {Ostriker}}{{Ricotti} \&
  {Ostriker}}{2004}]{ricotti04}
{Ricotti} M.,  {Ostriker} J.~P.,  2004, MNRAS, 352, 547

\bibitem[\protect\citeauthoryear{{Shlosman}, {Begelman} \& {Frank}}{{Shlosman}
  et~al.}{1990}]{shlosman90}
{Shlosman} I.,  {Begelman} M.~C.,    {Frank} J.,  1990, Nature, 345, 679

\bibitem[\protect\citeauthoryear{{Tremaine}, {Gebhardt}, {Bender}, {Bower},
  {Dressler}, {Faber}, {Filippenko}, {Green}, {Grillmair}, {Ho}, {Kormendy},
  {Lauer}, {Magorrian}, {Pinkney} \& {Richstone}}{{Tremaine}
  et~al.}{2002}]{tremaine02}
{Tremaine} S.,  {Gebhardt} K.,  {Bender} R.,  {Bower} G.,  {Dressler} A.,
  {Faber} S.~M.,  {Filippenko} A.~V.,  {Green} R.,  {Grillmair} C.,  {Ho}
  L.~C.,  {Kormendy} J.,  {Lauer} T.~R.,  {Magorrian} J.,  {Pinkney} J.,
  {Richstone} D.,  2002, ApJ, 574, 740

\bibitem[\protect\citeauthoryear{{Umemura}, {Loeb} \& {Turner}}{{Umemura}
  et~al.}{1993}]{umemura93}
{Umemura} M.,  {Loeb} A.,    {Turner} E.~L.,  1993, ApJ, 419, 459

\bibitem[\protect\citeauthoryear{{Volonteri}, {Haardt} \& {Madau}}{{Volonteri}
  et~al.}{2003}]{volonteri03}
{Volonteri} M.,  {Haardt} F.,    {Madau} P.,  2003, ApJ, 582, 559

\bibitem[\protect\citeauthoryear{{Volonteri} \& {Rees}}{{Volonteri} \&
  {Rees}}{2005}]{volonteri05b}
{Volonteri} M.,  {Rees} M.~J.,  2005, ApJ, 633, 624

\bibitem[\protect\citeauthoryear{{Warren}, {Quinn}, {Salmon} \&
  {Zurek}}{{Warren} et~al.}{1992}]{warren92}
{Warren} M.~S.,  {Quinn} P.~J.,  {Salmon} J.~K.,    {Zurek} W.~H.,  1992, ApJ,
  399, 405

\bibitem[\protect\citeauthoryear{{White} \& {Rees}}{{White} \&
  {Rees}}{1978}]{white78}
{White} S.~D.~M.,  {Rees} M.~J.,  1978, MNRAS, 183, 341

\end{thebibliography}

\begin{figure*}
\centerline{\epsfig{figure=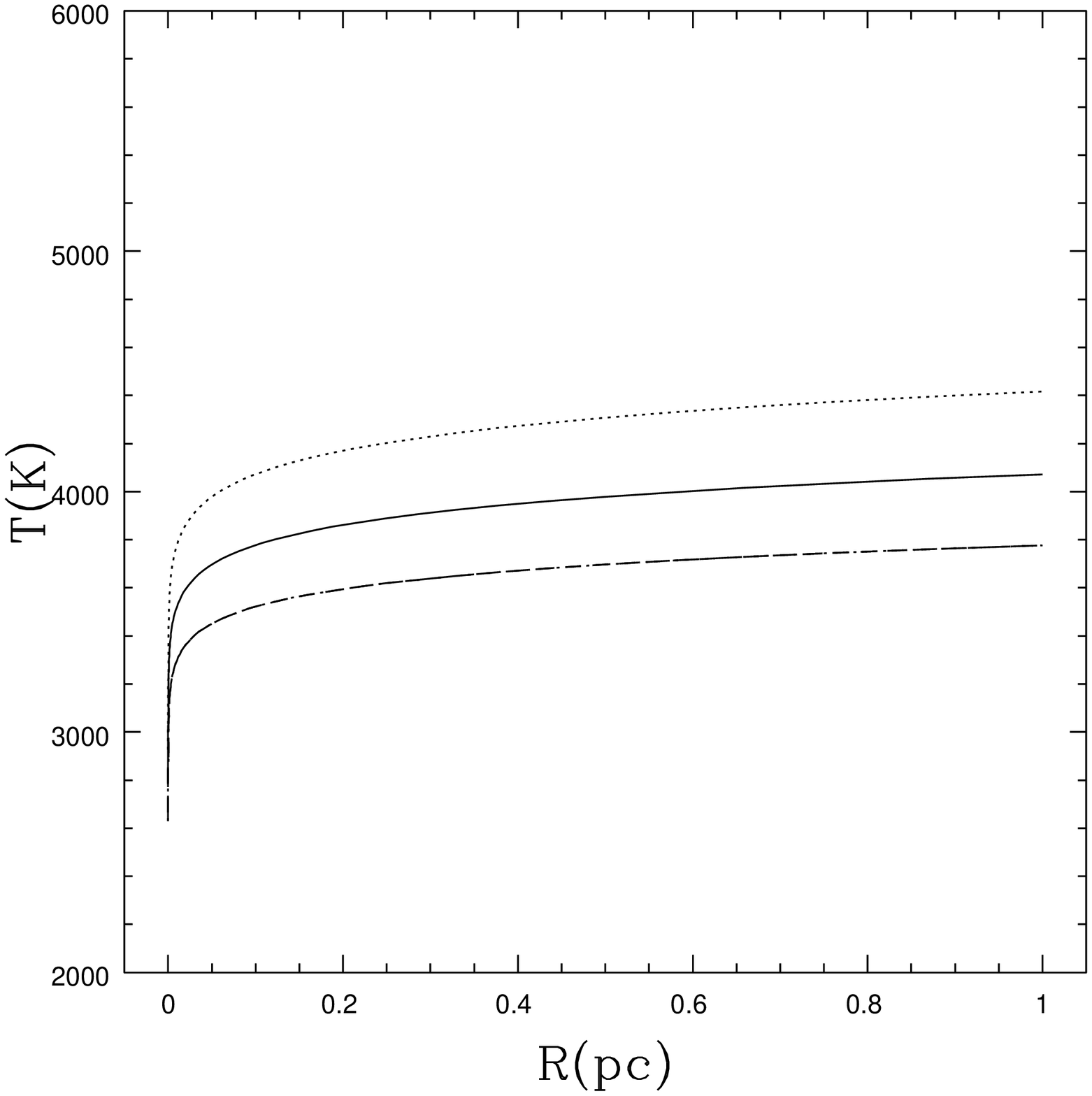,width=0.4\textwidth}
            \epsfig{figure=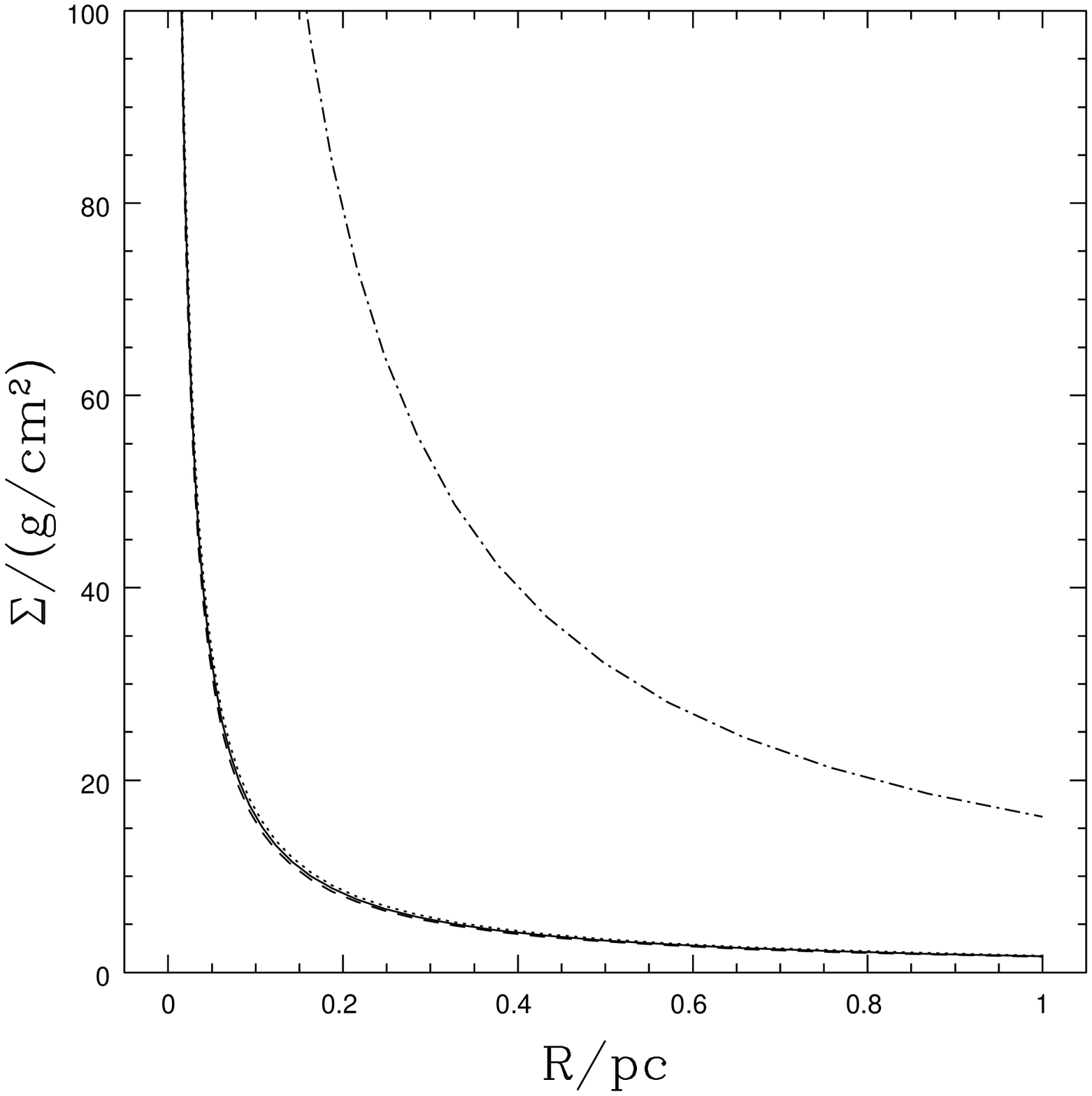,width=0.4\textwidth}} 
\centerline{\epsfig{figure=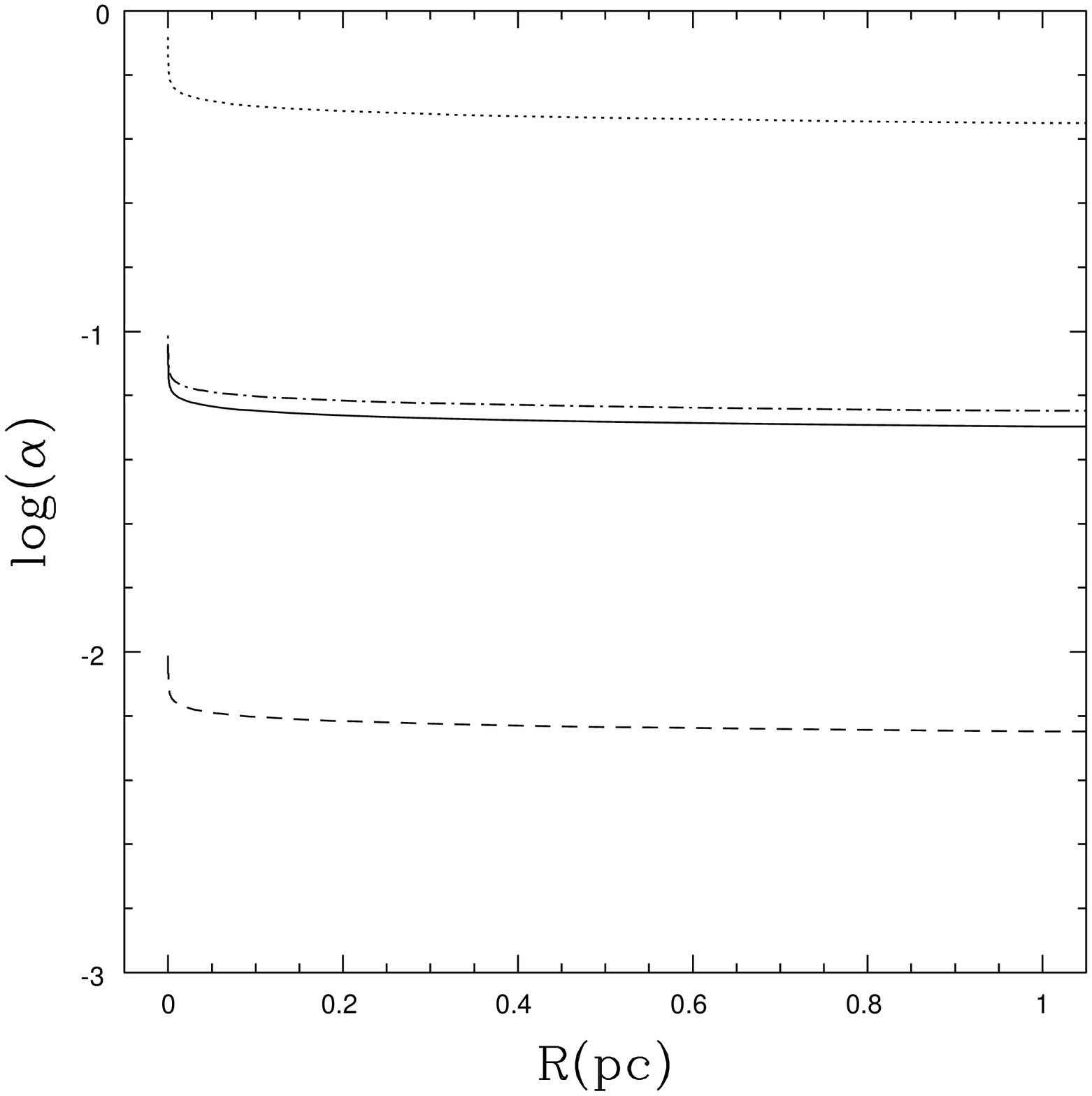,width=0.4\textwidth}
            \epsfig{figure=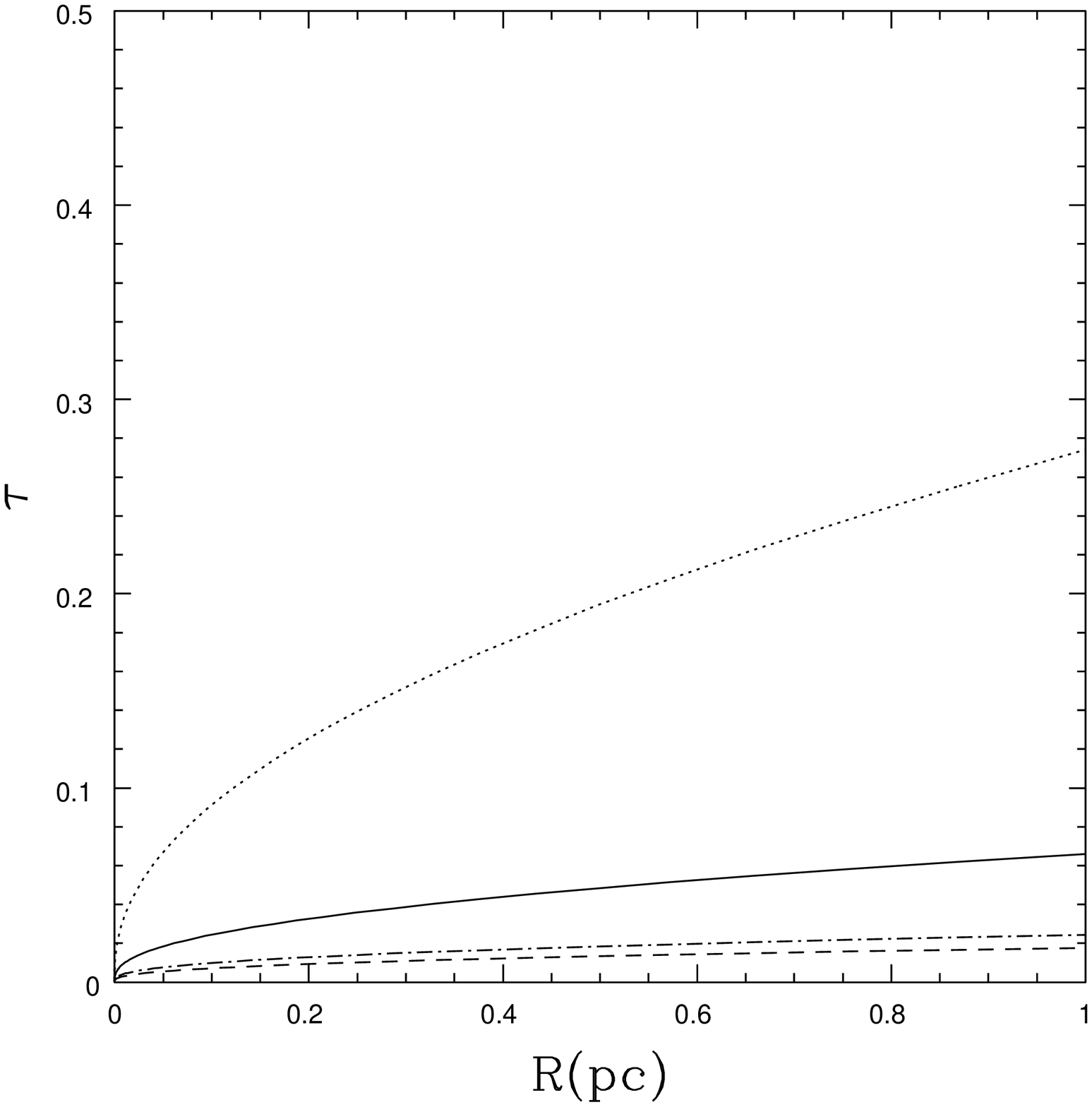,width=0.4\textwidth}}
\caption{Structure of self-regulated accretion discs with cooling
dominated by atomic hydrogen. The plots show (clockwise from top left):
the temperature $T_{\rm gas}$, the surface density $\Sigma$, the
optical depth $\tau$ and the viscosity coefficient $\alpha$ . The lines
refer to the following pairs of $(\mdot,V_{\rm h})$: solid line
($10^{-2},10$), dotted line ($10^{-1},10$), dashed line ($10^{-3},10$),
dot-dashed line ($10^{-2},100$), where $\mdot$ is in units of $\msunyr$ and
$V_{\rm h}$ in km/sec.}
\label{fig:structure}
\end{figure*}

\section*{Appendix: The structure and stability of self-gravitating
zero-metallicity accretion discs}

We consider the structure and stability of self-gravitating discs
embedded in the external potential generated by a dark matter halo. If
the disc is cold enough (or massive enough) it can become
self-gravitating, i.e. it will develop gravitational instabilities. It
is well known \citep{gammie01,LR04,LR05} that once gravitational
instabilities set in, the disc will rapidly evolve into a quasi-steady
state of marginal gravitational instability, characterized by a
constant profile of the Toomre parameter $Q$.

We start by noting that the requirement that $Q$ be constant, for our
flat rotation curve disc, implies that if the disc is isothermal (as it
is expected to be, see \citealt{oh} and arguments below) then the
surface density $\Sigma\propto R^{-1}$, i.e. the disc has the same
density profile as a simple Mestel disc. This in principle, makes it
very easy to include the contribution of the disc itself to the radial
gravitational field, since it also leads self-consistently to a flat
rotation curve. However, we will not consider this contribution at
present. The steady state structure of such Mestel radially
self-gravitating accretion discs has been explored extensively by
\citet{BL99}, who also include the modifications to the disc structure
expected to take place very close to the black hole, where its
gravitational field starts to become important.

\begin{figure*}
\centerline{\epsfig{figure=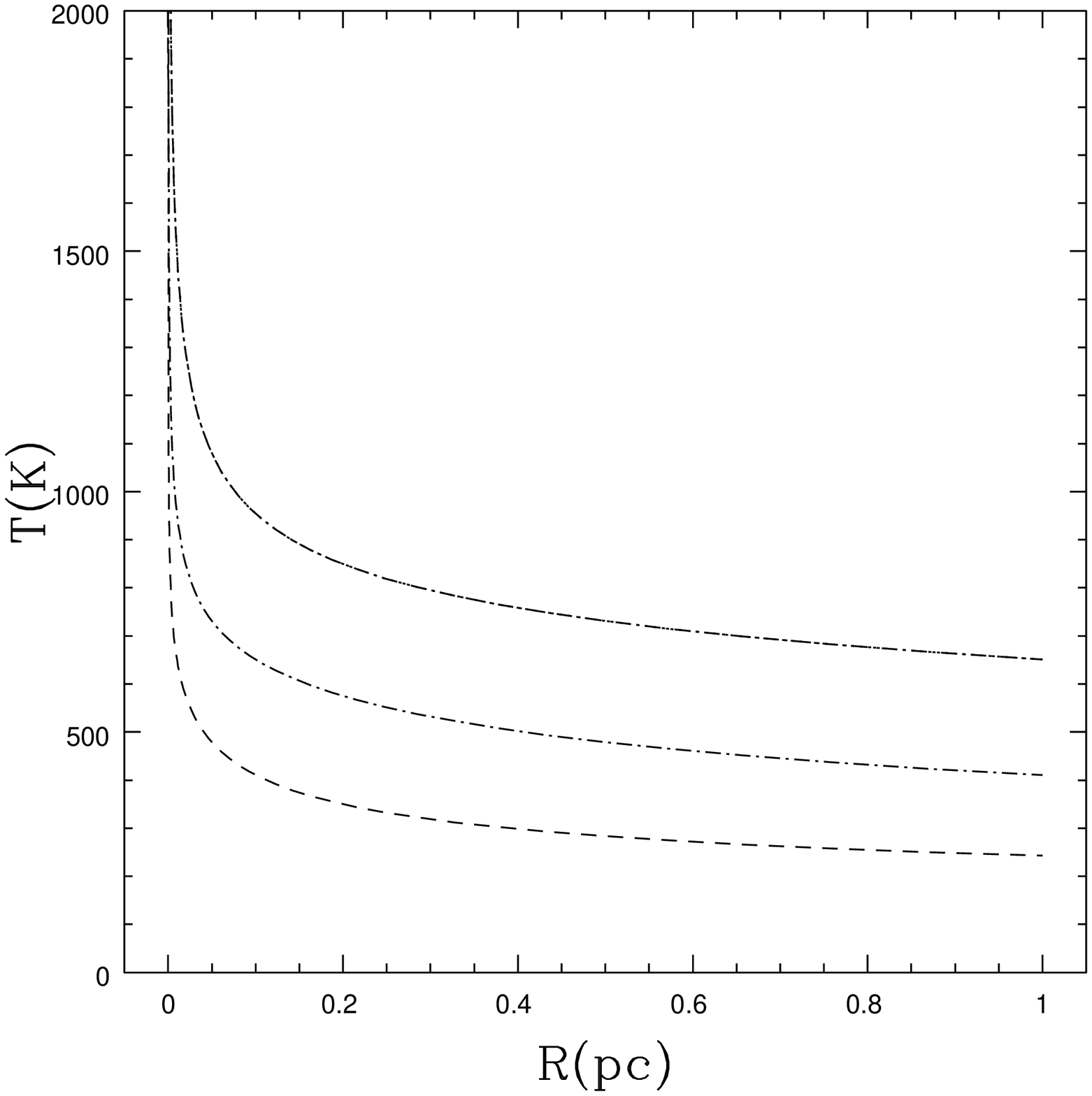,width=0.3\textwidth}
            \epsfig{figure=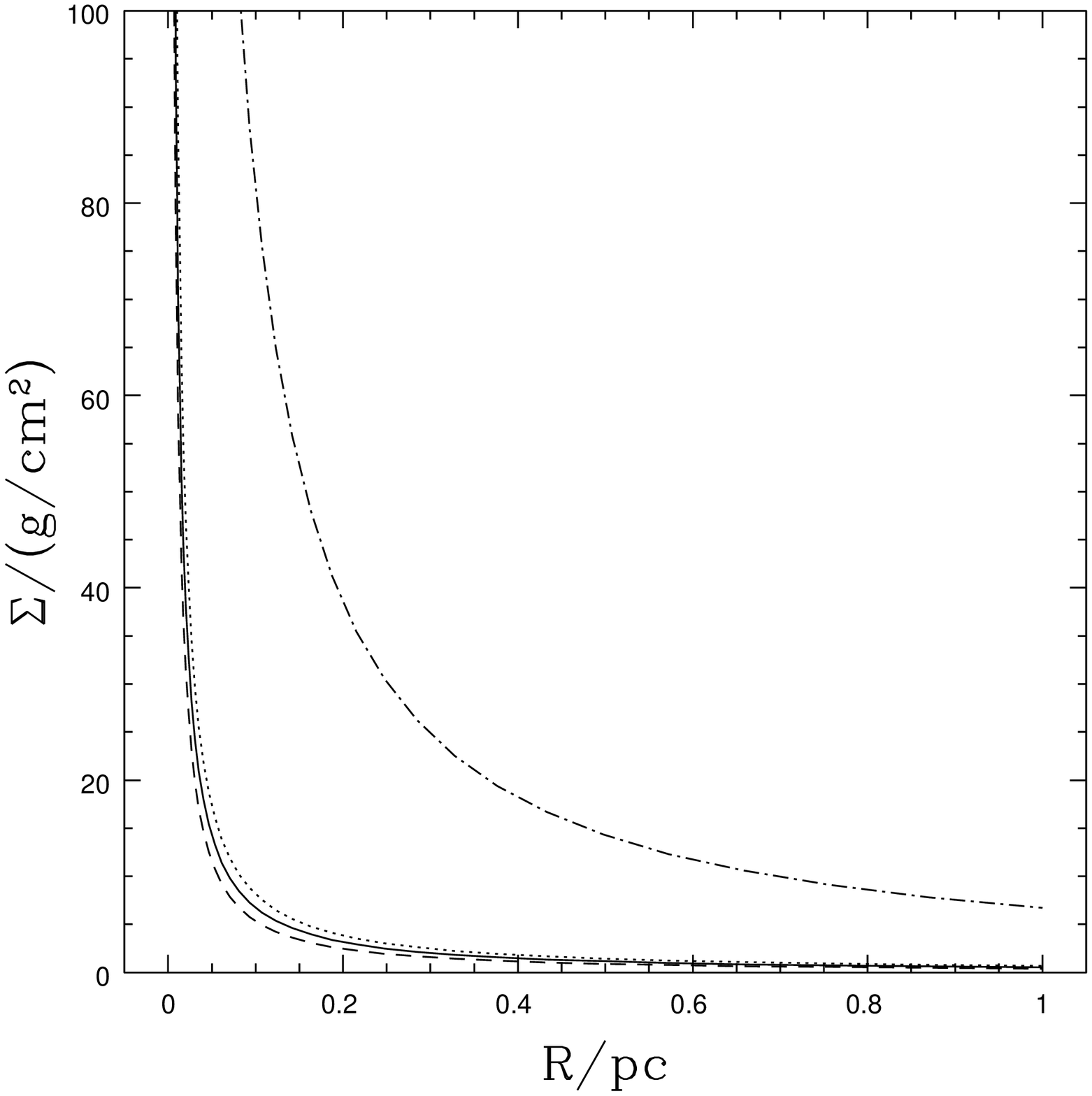,width=0.3\textwidth} 
            \epsfig{figure=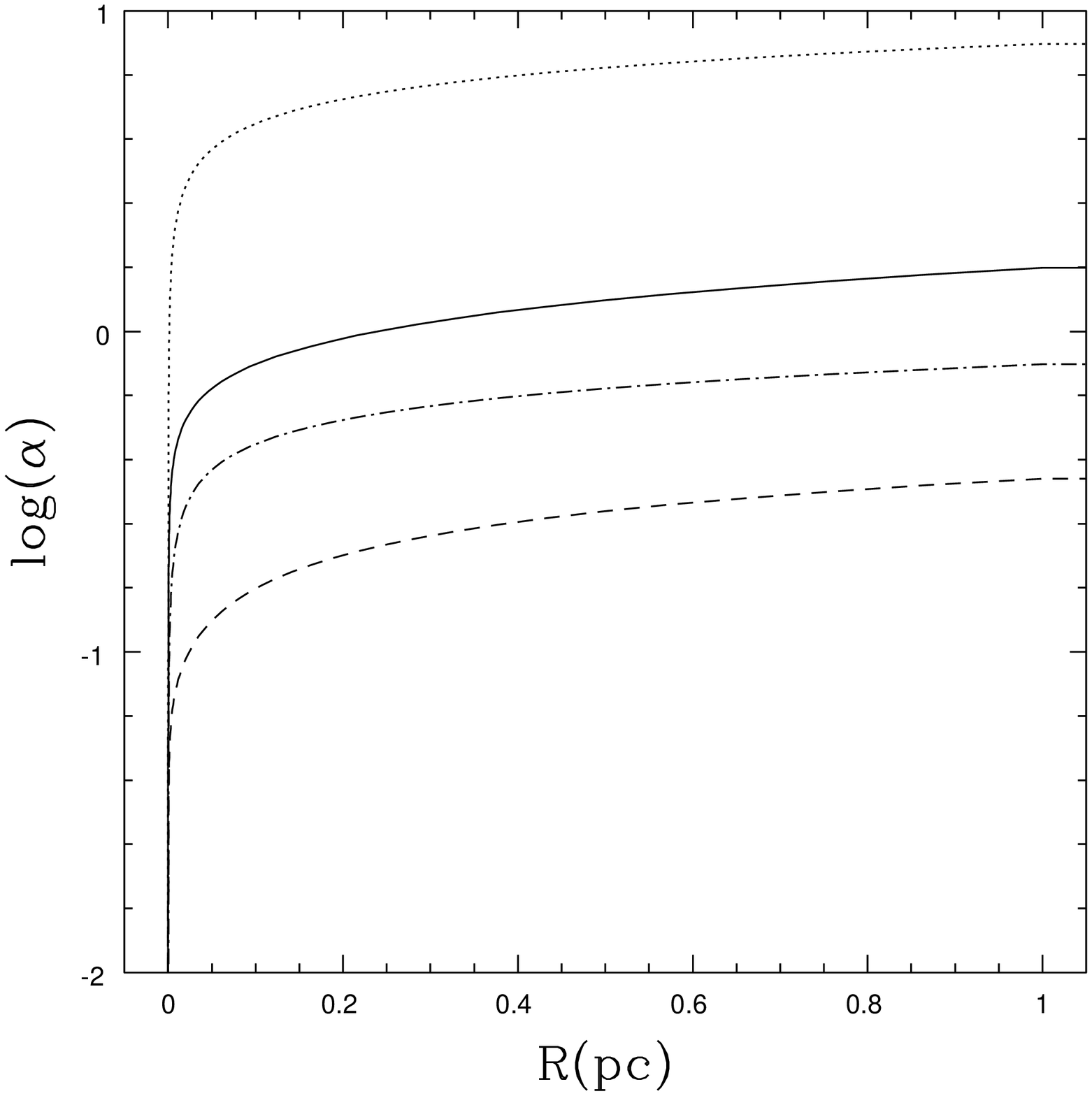,width=0.3\textwidth}}
\caption{Structure of self-regulated accretion discs with cooling
dominated by molecular hydrogen. The plots show (clockwise from top
left): the temperature $T_{\rm gas}$, the surface density $\Sigma$, and
the viscosity coefficient $\alpha$. The lines refer to the following
pairs of $(\mdot,V_{\rm h})$: solid line ($10^{-2},10$), dotted line
($10^{-1},10$), dashed line ($10^{-3},10$), dot-dashed line ($10^{-2},100$),
where $\mdot$ is in units of $\msunyr$ and $V_{\rm h}$ in km/sec.}
\label{fig:structure2}
\end{figure*}

Actually, it is possible to solve for the detailed disc structure by
relaxing the assumption of isothermality of the disc, and by computing
the temperature by requiring that radiative cooling (here assumed to be
optically thin) is balanced by the heating provided by the accretion
mechanism. This energy balance equation gives us a relationship between
$\alpha$ and the cooling time $t_{\rm cool}$ \citep{pringle81}:

\begin{equation}
\alpha=\frac{1}{\gamma(\gamma-1)}\frac{t_{\rm dyn}}{t_{\rm cool}},
\label{alpha}
\end{equation}
where $\gamma$ is the ratio of specific heats, $t_{\rm dyn}=R/V_{\rm
h}$ is the dynamical timescale, and $t_{\rm cool}=nkT_{\rm gas}/
n^2\Lambda(T_{\rm gas})$, where $n$ is the number density, $T_{\rm
gas}$ is the gas temperature and $\Lambda$ is the relevant cooling
function. 

Once the cooling function is provided, it is then possible to solve the
basic disc equations describing self-regulation (eqn. (\ref{Q})),
angular momentum conservation (eqn. (\ref{eq:mdot})) and thermal
equilibrium (eqn. (\ref{alpha})) for the three unknowns $\Sigma(R)$,
$T_{\rm gas}(R)$ and $\alpha(R)$, with the input parameters $Q$,
$\mdot$ and $V_{\rm h}$.

We consider here the case where the disc is made of primordial gas,
with no metal enrichment. The cooling process is then going to be
provided by hydrogen. We consider two cases: (a) when
molecular hydrogen formation is suppressed, so that
cooling is dominated by atomic hydrogen, and (b) the case where
molecular hydrogen is present and therefore dominates the cooling
function.

\subsection{Cooling processes: Atomic hydrogen cooling}

We consider here the cooling function for pure atomic hydrogen provided
by \citet{cen1992} (we have also considered a different cooling
function, after \citealt{katz}, and found no significant
difference). The results of the calculations are shown in
Fig. \ref{fig:structure} where we plot the resulting profiles of
$\Sigma$, $T_{\rm gas}$ and $\alpha$ for different values of the input
parameters. Here, we have assumed $Q=1$ and varied $\mdot$ and $V_{\rm
h}$. As can be seen, due to the steepness of the cooling function, the
disc turns out to be almost isothermal, with the equilibrium
temperature only very weakly dependent both on radius and especially on
input parameters, with $T_{\rm gas}\approx 4000$ K (note that changing
the mass accretion rate by two orders of magnitude, only modifies
$T_{\rm gas}$ by roughly 14\%). In Fig. \ref{fig:structure}, we also
plot the optical depth of the disc $\tau=\Sigma\kappa_{\rm R}$, where
the Rosseland mean opacities $\kappa_{\rm R}$ for a primordial
composition has been taken from \citet{mayer05}. We see that the
assumption of optically thin cooling is indeed valid over the whole
disc.

\subsection{Cooling processes: Molecular hydrogen cooling}

We consider next the situation wherein molecular hydrogen formation is
not suppressed and we assume that it can form up to a molecular
fraction $x_{H_2}=10^{-3}$ \citep{oh}. The cooling function for
molecular hydrogen is taken from
\citet{galli98}. Fig. \ref{fig:structure2} shows $\Sigma$, $T_{\rm
gas}$ and $\alpha$ for this case. We do not plot the optical depth for
molecular hydrogen cooling since the available opacity tables for
primordial composition do not cover the very low temperature range
predicted. At any rate, the opacity is expected to be extremely low. We
can see that the equilibrium temperature is going to be much lower than
in the case of atomic hydrogen cooling, as expected.

Since the cooling function is less strongly dependent on temperature
for molecular hydrogen cooling, the ``thermostat'' that keeps the
temperature of the disc constant in the case of atomic cooling is less
efficient, and the equilibrium temperature in this case depends more
strongly on the input parameters. The equilibrium temperatures range
from 200 K to 700 K. The disc is still close to isothermal, with the
temperature rising in the inner disc.

\end{document}